# A Hardware and Software Platform for Aerial Object Localization

Matthew Szenher[*,**], Alex Delacroix[†,††], Eric Keto[‡,‡‡], Sarah Little[§], Mitch Randall[¶], Wesley Andres Watters[‖], Eric Masson[*] and Richard Cloete[†]

[*]Galileo Project, Center for Astrophysics
60 Garden Street, Cambridge, MA 02139, USA

[†]Center for Astrophysics, Harvard & Smithsonian
Cambridge, MA 02138, USA

[‡]Institute for Theory and Computation
Harvard College Observatory, Cambridge, MA 02138, USA

[§]Scientific Coalition for UAP Studies, Wellesley, MA 02138, USA

[¶]Ascendant AI, Boulder, CO 80304, USA

[‖]Whitin Observatory, Wellesley College
20 106 Central St. 21 Wellesley, MA 02481, USA

[**]mszenher@yahoo.com
[††]Alex.delacroix@cfa.Harvard.edu
[‡‡]eketo@cfa.harvard.edu

To date, there are little reliable data on the position, velocity and acceleration characteristics of Unidentified Aerial Phenomena (UAP). The dual hardware and software system described in this document provides a means to address this gap. We describe a weatherized multi-camera system which can capture images in the visible, infrared and near infrared wavelengths. We then describe the software we will use to calibrate the cameras and to robustly localize objects-of-interest in three dimensions. We show how object localizations captured over time will be used to compute the velocity and acceleration of airborne objects.

## 1. Introduction

The work of the Galileo Project pertains to the detection and characterization of Unidentified Aerial Phenomena (UAP) for the purpose of their scientific study. In particular, we are interested in collecting data that can distinguish between human-made objects, naturally occurring effects, and phenomena that do not fit in either category (Watters, 2023).

Here, we focus on a suite of cameras which we will use to infer the kinematics of objects in the sky from their tracks. A track is a time-series of object localizations in three-dimensional (3D) space. A single imaging sensor is capable of obtaining a two-dimensional (2D) projection onto an image plane of the 3D location of an aerial object. At least two cameras spaced apart by a known distance and pointing in known directions (i.e. calibrated cameras) are required to resolve the position of an object in 3D space (Hartley & Zisserman, 2004). The object must be visible to both cameras at the same time, and these must sample at nearly the same time in order to resolve the instantaneous location of an object that is moving. Further, multiple samples must be taken sequentially at known times in order to form the complete 3D track and reconstruct the full kinematics of the object in the sky. There is an important scientific utility in determining object tracks in 3D — as well as inferring the velocities and accelerations (Knuth *et al.*, 2019) of these objects — as these data could potentially be used to distinguish anomalous objects from prosaic objects as described in Watters (2023).

[**]Corresponding author.

We have chosen to use optical, infrared and near infrared cameras — among others — since UAP reports in the media (Cooper *et al.*, 2017), in the "gray literature" (Watters, 2023), by the US government (Office Of The Director Of National Intelligence, 2021) and in the scientific literature (Knuth *et al.*, 2019; Hernandez *et al.*, 2018) indicate that UAPs may be visible at these wavelengths. Also, infrared data can be used to detect and characterize exhaust heat (or the lack thereof) from tracked objects. An additional advantage of long-wave infrared cameras is their exceptional detection performance under a range of conditions and across larger distances. This is largely because detection depends primarily on object surface emission rather than reflection. Thermal infrared cameras are able to detect objects more readily at night, and detection does not depend on illumination geometry (which can severely compromise the detectability at optical wavelengths of objects that are visible primarily because they reflect sunlight).

This paper makes three novel contributions as follows:

- We describe a unique suite of cameras used to collect data for the scientific study of UAP (Sec. 3).
- We introduce a new algorithm to calibrate cameras using data from Automatic Dependent Surveillance-Broadcast (ADS-B)-equipped aircraft (Sec. 4.2.1).
- We describe a novel event-driven algorithm to collect data from remote cameras for the purposes of object localization (Sec. 4.5).

This paper is organized as follows. In Sec. 2, we describe in detail the basic problem to be solved: using simultaneous frames from multiple cameras to find the instantaneous location of an object in space. In Sec. 3, we describe the cameras (visual, infrared, and near infrared) which comprise our imaging suite. This section also describes the way we intend to protect these outdoor cameras against sun, heat, wind, and weather. Section 4 outlines the software we have written to solve the object tracking problem using the cameras described in Sec. 3. In particular, Secs. 4.1 and 4.2 detail how we will calibrate our cameras. Section 4.5 describes our object localization algorithm. We describe a number of simulated experiments used to assess this localization algorithm in Sec. 5. In Sec. 5, we also provide the results of these experiments with related discussion. Section 6 describes relevant work from meteor and fireball networks and observatories. We close with conclusions and future work in Sec. 7.

## 2. Background

### 2.1. *Definition of the Object Localization Problem with Multiple Cameras*

A single camera suffices only to provide an imaged object's angular position with respect to that camera's pose (i.e. location and orientation), such as azimuth and elevation but not range. An object must be imaged in multiple cameras simultaneously to infer that object's 3D location with respect to a coordinate system in which the poses of those cameras are expressed. That object must be imaged over multiple image frames — and recognized as the same object in each frame–in order for its track to be computed and its velocity and acceleration inferred.

We introduce three related coordinate systems which will be referred to throughout this work: the image coordinate system, the camera coordinate system, and the world coordinate system. Details on these coordinate systems are given in the caption of Fig. 1. All coordinate systems are right-handed in this work.

A camera's intrinsic parameters (also known as internal parameters) are used to relate a point in camera coordinates to image coordinates (Hartley & Zisserman, 2004). The intrinsic parameters are

- the camera's focal length expressed in pixel units in the $x$ and $y$ directions on the image plane: $(\alpha_x, \alpha_y)$;
- the camera's optical center $(c_x, c_y)$;
- and a set of distortion coefficients which describe how the camera's lens refracts incoming light rays.

Note that image plane skew is sometimes included as an intrinsic parameter. We assume skew is negligible and therefore ignore it in this paper.

As described in Sec. 4, a camera's distortion coefficients are used to convert an image point $(x_i, y_i)$ captured by a lensed camera to the equivalent image point $(x'_i, y'_i)$ had that lens been removed (i.e. had the camera been a pinhole camera). We will call this conversion process "undistortion". The relationship between a point $(X_c,\ Y_c,\ Z_c)$ in camera coordinates to an undistorted image point $(x_i',\ y_i')$ (as a homogeneous vector) is given by

$$s\begin{pmatrix} x_i' \\ y_i' \\ 1 \end{pmatrix} = \begin{pmatrix} \alpha_x & 0 & c_x \\ 0 & \alpha_y & c_y \\ 0 & 0 & 1 \end{pmatrix}\begin{pmatrix} X_c \\ Y_c \\ Z_c \end{pmatrix}. \tag{1}$$

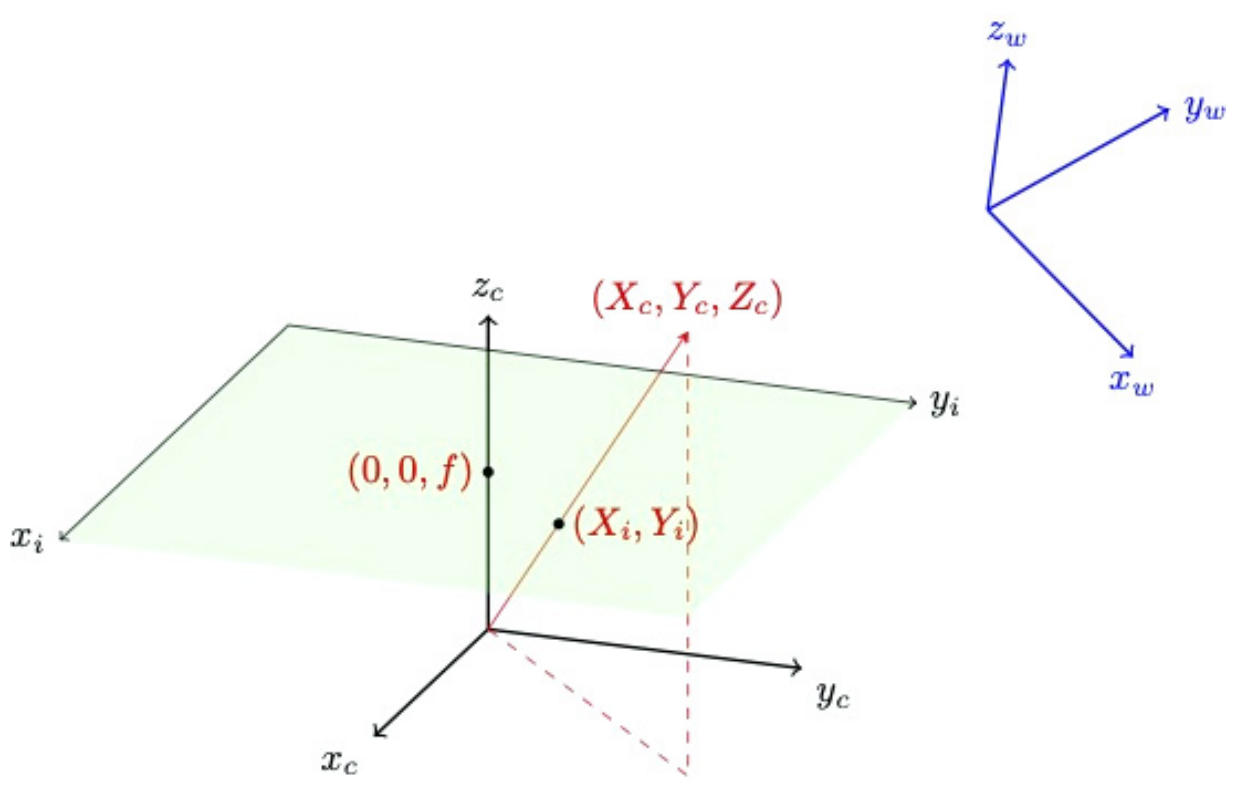

Fig. 1. (Color online) A single camera's coordinate system is represented by the coordinate system with axes $x_c$, $y_c$, $z_c$. We assign the $z_c$ axis to be the camera's optical axis; it is orthogonal to the camera's image plane (green rectangle). The value $f$ is the camera's focal length. A point at $(X_c,\ Y_c,\ Z_c)$ in camera coordinates will be imaged at $(X_i,\ Y_i)$ on the image plane (ignoring for the moment any refraction due to camera lenses). We define the image plane's $x$-axis ($x_i$) to be parallel to the camera's $x$-axis ($x_c$) and the same goes for the $y$-axes. The world coordinate system is represented as the blue axes in this figure. The camera local coordinate system is related to the world coordinate system by a rotation and translation operation.

These intrinsic parameters reflect a camera's internal geometry and generally need only be computed once per camera (or unless the internal geometry changes due to e.g. a mechanical zoom which changes the focal length of the camera). Section 4 describes how we will compute our cameras' intrinsic parameters. We use pixel units to represent $\alpha_x$, $\alpha_y$, $c_x$ and $c_y$ in this paper. The matrix of intrinsic parameters in Eq. (1) will be labeled $K$ in the rest of this paper.

While we can use Eq. (1) to project from 3D camera coordinates to 2D image coordinates, we generally cannot recover the 3D camera coordinates of an imaged object using 2D image coordinates from a single camera. We can, however, use 2D image coordinates to compute the direction of the object in the camera coordinate frame. We will call this direction vector a "look-at" vector in the rest of this work.

A camera's extrinsic parameters (also known as its external parameters) are used to relate a 3D point in homogeneous world coordinates to that same point's 3D representation in homogeneous camera coordinates (Hartley & Zisserman, 2004). The extrinsic parameters encode the rotation and translation between the camera and world coordinate systems. Section 4 describes how we will compute our cameras' extrinsic parameters.

Together, a camera's intrinsic matrix and the external parameters will be used to form a projection matrix $P$ which is used to map a 3D point in world coordinates $X$ to an undistorted 2D point $x_i$ in image coordinates:

$$x_i = PX. \tag{2}$$

We define a calibrated camera as one whose intrinsic and extrinsic parameters are known.

In order to estimate the location of the centroid of an imaged object in the world coordinate system, that object must be imaged nearly simultaneously by at least two calibrated cameras; note that acquiring nearly simultaneous images taken by two different camera image streams is not a trivial problem, and we describe our solution to it in Sec. 4.4. We must then infer which detected object centroid point in the first image corresponds to a given detected object centroid point in the second image. A point in the first image corresponds to a point in the second image if and only if they represent the same object (Bach and Aggarwal, 1988). We can then use Eq. (1) to convert our image points to look-at unit vectors in camera coordinates, and follow with an application of the extrinsic parameters to convert those look-at vectors to world coordinates. When there are no errors in camera calibration, image synchronization and object correspondence, the look-at vectors in world coordinates from each camera origin to the object's centroid will intersect perfectly at the object's centroid in 3D world coordinates. It is then straightforward to compute the object's 3D centroid in world coordinates by converting each look-at vector to a half-infinite ray beginning at each camera's origin (look-at rays) and solving for the intersection of any pair of those rays. Unfortunately, the situation is rarely that straightforward, as we discuss in Sec. 2.2.

Cloete (2023) describes our object detection algorithm, based on the You Only Look Once (YOLO) framework. We use YOLO for detecting objects of interest in images (required for object localization). The Deep Simple Online and Realtime Tracking (DeepSORT) algorithm (Cloete, 2023) is used to link these detections together across multiple frames (required for object tracking and object velocity and acceleration estimation). The object detection algorithm is run on each camera's image stream and returns the image coordinates of the centroid of a detected object of interest as input to our localization algorithm.

### 2.2. *General Triangulation Error Analysis*

It is helpful to derive a generalized error analysis to be used in gauging instrument performance based on parameters such as baseline length, object distance, and object angles with respect to the observers. In order to generalize, we assume a two-camera system (stereo). These two cameras ($L$ and $R$) and an object ($O$) define the plane shown in Fig. 2.

We define $Z$ to be the distance from camera $L$ to object $O$. We compute the distance starting with the law of sines

$$\frac{Z}{\sin\beta} = \frac{b}{\sin\gamma}. \tag{3}$$

We substitute $\gamma = \pi - \theta - \beta = \alpha - \beta$ and solve for $Z$

$$Z = \frac{b\sin\beta}{\sin(\alpha - \beta)}. \tag{4}$$

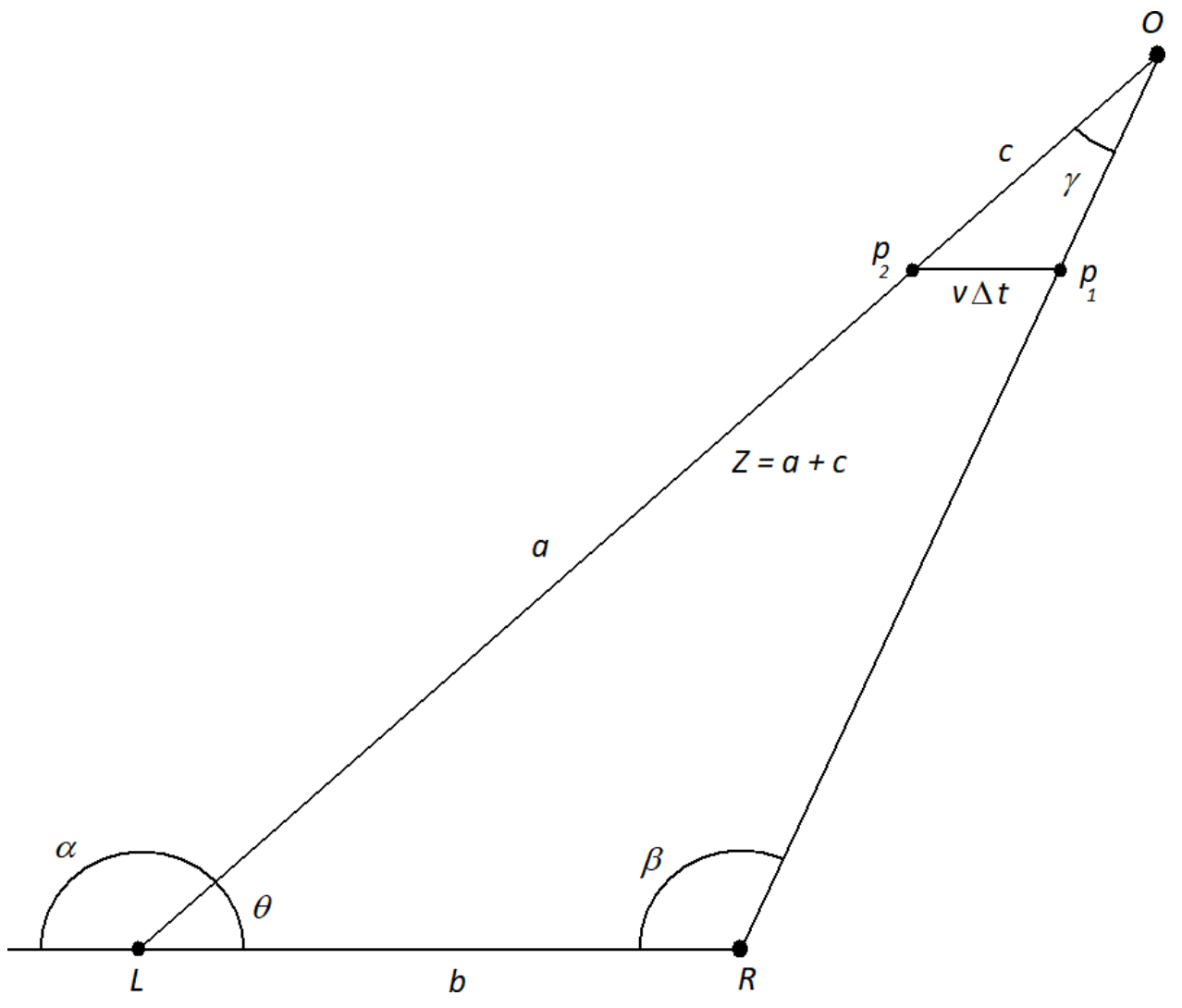

Fig. 2. Two cameras labeled $L$ and $R$ form a baseline of length $b$ and point with angles $\alpha$ and $\beta$ with respect to the baseline. Their lines of sight (i.e. their look-at vectors) intersect at point $O$ a distance $Z$ from camera $L$. In time $\Delta t$, an object of velocity $v$ moves parallel to the baseline from point $p_1$ to $p_2$.

To compute the error in distance $Z$ due to the error in angle $\beta$ we take the partial derivative of $Z$ with respect to $\beta$ and use Eq. (4) to simplify terms

$$\frac{\partial Z}{\partial\beta} = Z\frac{\sin\alpha}{\sin\beta}\frac{1}{\sin(\alpha-\beta)}. \tag{5}$$

When $\alpha$ and $\beta$ are nearly equal, we can approximate $\sin(\alpha-\beta)$ with $(\alpha-\beta)$ and $\frac{\sin\alpha}{\sin\beta}$ with 1 so in this case Equation (5) becomes dominated by $\frac{1}{\alpha-\beta}$ and this term becomes very large. Thus, the accuracy becomes very poor when the two angles are nearly equal. Manipulating the partial derivative slightly, we see that

$$Z\frac{\sin\alpha}{\sin\beta}\frac{1}{\sin(\alpha-\beta)} = Z^2\frac{\sin\alpha}{b\sin^2\beta}. \tag{6}$$

Thus, we see that the measurement error in the object distance, $Z$, due to angle uncertainty, depends on the square of the distance, $Z^2$, and is inversely proportional to the baseline distance, $b$.

We now investigate errors due to a lack of time synchronization between cameras $L$ and $R$.

Suppose that the cameras' captured images are separated by time $\Delta t$. Suppose also that a moving object travels parallel to the baseline from $p_1$ to $p_2$ a distance $v\Delta t$ in that time, where $v$ is the magnitude of the object's velocity vector. In this case, the cameras will measure an apparent distance $Z$ whereas the actual distance was $a$. We can compute this error by noting that the line segment from $p_1$ to $p_2$ is parallel with the baseline so that two similar triangles are formed, thus

$$\frac{v\Delta t}{b} = \frac{c}{a+c}. \tag{7}$$

Solving for the fractional error, we write

$$\frac{c}{a} = \frac{v\Delta t}{b - v\Delta t}. \tag{8}$$

This result tells us that as the distance traveled $v\Delta t$ between images grows to the size of the baseline, large errors in distance estimation result. The above error analysis focuses on the case in which the object's velocity vector is parallel to the camera baseline. We explore the general case — in which the velocity vector is directionally unconstrained — in Appendix B.

## 3. Hardware Implementation

The primary suite of cameras used to detect and track objects will monitor the entire sky in the infrared (IR), near infrared (NIR), and optical wavelengths. As depicted in Fig. 3, this suite includes IR and optical/NIR cameras, code-named "the Dalek", which is a custom arrangement of eight FLIR Boson 640 and one ZWO 462MC visible/NIR wide-field cameras; and a stand-alone, visible all-sky camera: an Alcor OMEA 9C. We describe each of these cameras systems in detail below.

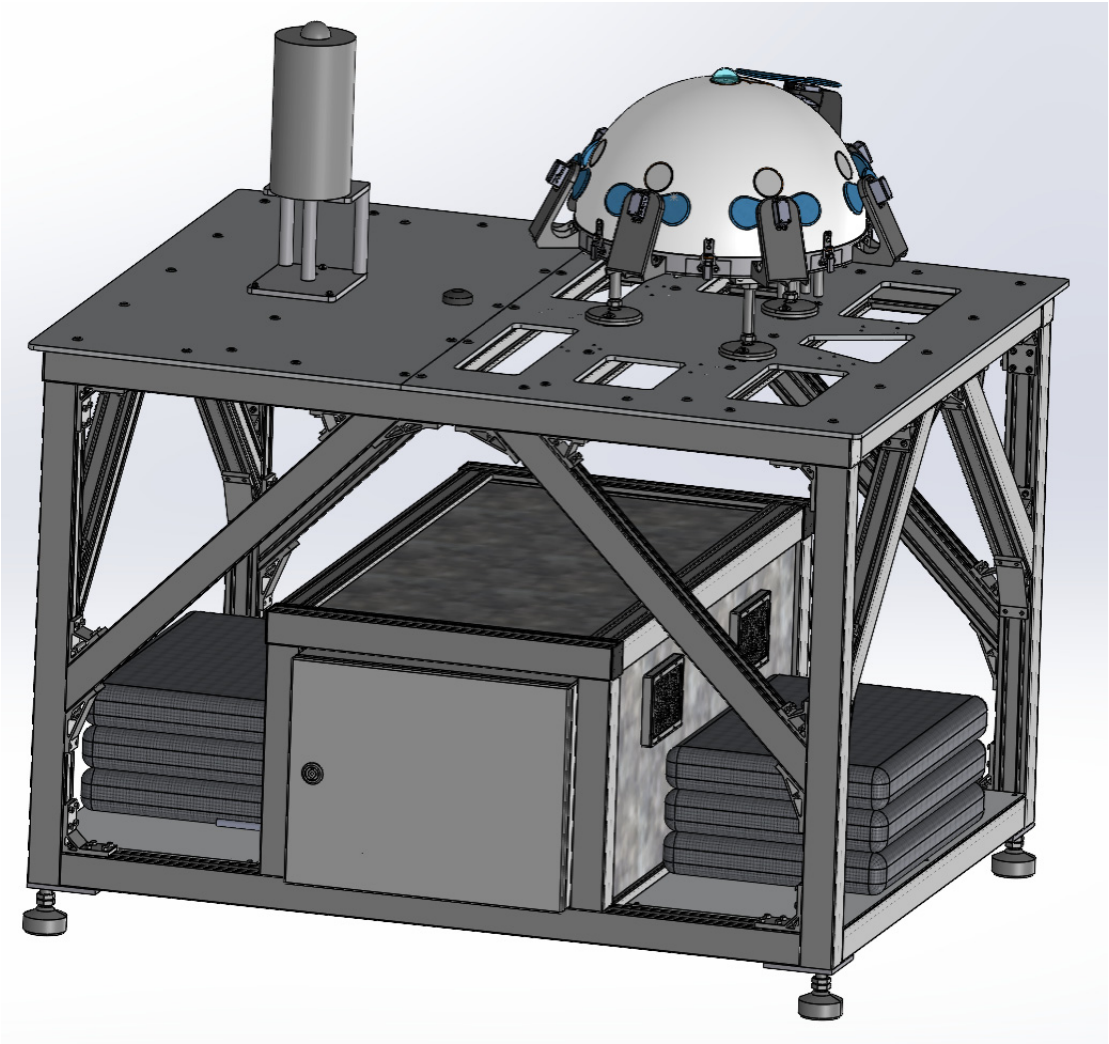

Fig. 3. Alcor visible all-sky camera tower; and the Dalek dome with IR camera array and all-sky visible/NIR camera. The cabinet underneath contains a computer which is responsible for running object detection as well as transmitting data to a central repository. The cabinet is an electromagnetic interference (EMI)/radio frequency interference (RFI) shielded, ingress protection rated (IP65), weatherproofed enclosure.

### 3.1. *The Dalek Infrared and Near Infrared Camera Array*

As pictured in Fig. 4(a), the Dalek is a purpose-built hemispherical array of eight IR and one zenith all-sky NIR camera. These cameras are installed in a fiberglass dome enclosure and faced with sealed, Germanium windows. Seven FLIR Boson 640 long-wave IR (LWIR) (7.5 $\mu$m to 13.5 $\mu$m) cameras (see Table 1 for full specifications) are arranged radially to provide a 360° view of the area; these cameras each have a 50° FOV and are nominally pointing 30° above the horizon (see Fig. 4(b)). The ability of IR radiation to penetrate the atmosphere without being absorbed by water, oxygen, or carbon dioxide molecules depends on its wavelength. The Boson LWIR cameras detect IR light in the atmosphere's high transmittance window between 8 $\mu$m and 14 $\mu$m (see Fig. 5).

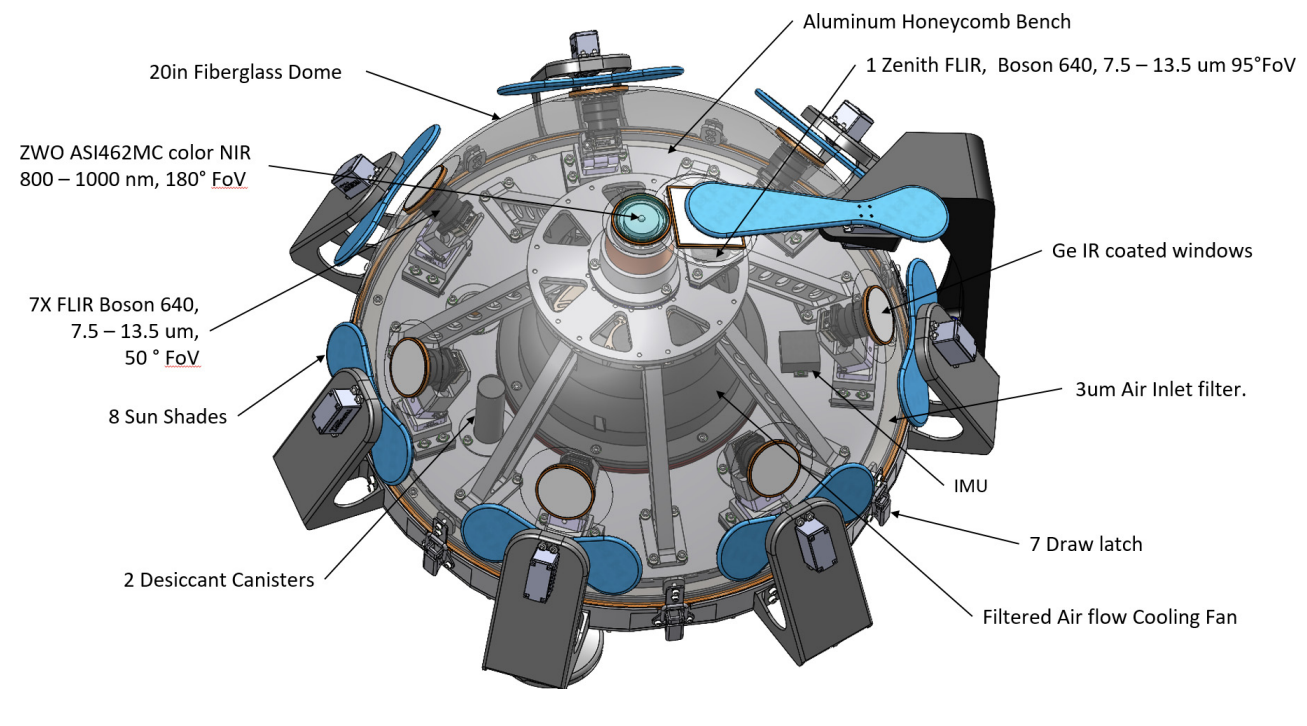

(a)

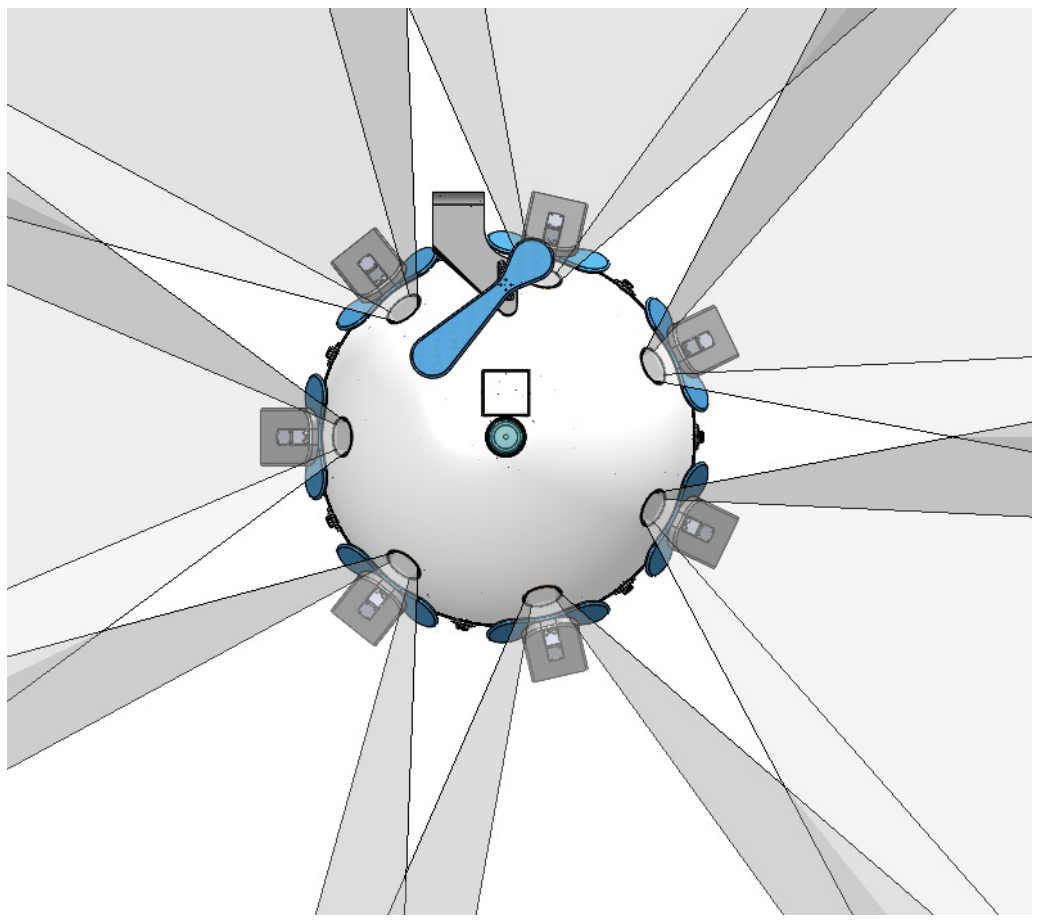

(b)

Fig. 4. (a) The infrared camera array (Dalek) is an array of eight IR and one zenith all-sky visible/NIR camera. Seven IR cameras are arranged hemispherically to provide a 360° view of the surroundings. One IR and one NIR camera point towards the zenith. These cameras are installed in a fiberglass dome enclosure and protected from the elements with Germanium windows. (b) The field of view (FOV) of the Dalek's seven radial IR cameras is 50°. Adjacent cameras' FOVs overlap.

Our preliminary tests indicate that large airliners (60 m long) are easily detectable by an FLIR Boson 640 camera at a range of 20 km and barely detectable at a range of 35 km. A 20-m long business jet is likewise barely detectable at a range of 23 km,

Table 1. FLIR Boson 640 camera specifications.

| Thermal Imaging Detector | Uncooled VOx microbolometer |
|---|---|
| Field of View | 7 radial: 50°; 1 zenith: 95° |
| Resolution | 640 × 512 pixels |
| Pixel size | 2.9 $\mu$m × 2.9 $\mu$m |
| Frame rate | 60 Hz baseline |
| | 30 Hz runtime selectable |
| Thermal Spectral range | 7.5 $\mu$m – 13.5 $\mu$m |
| Working temperature | −40°C–80°C |

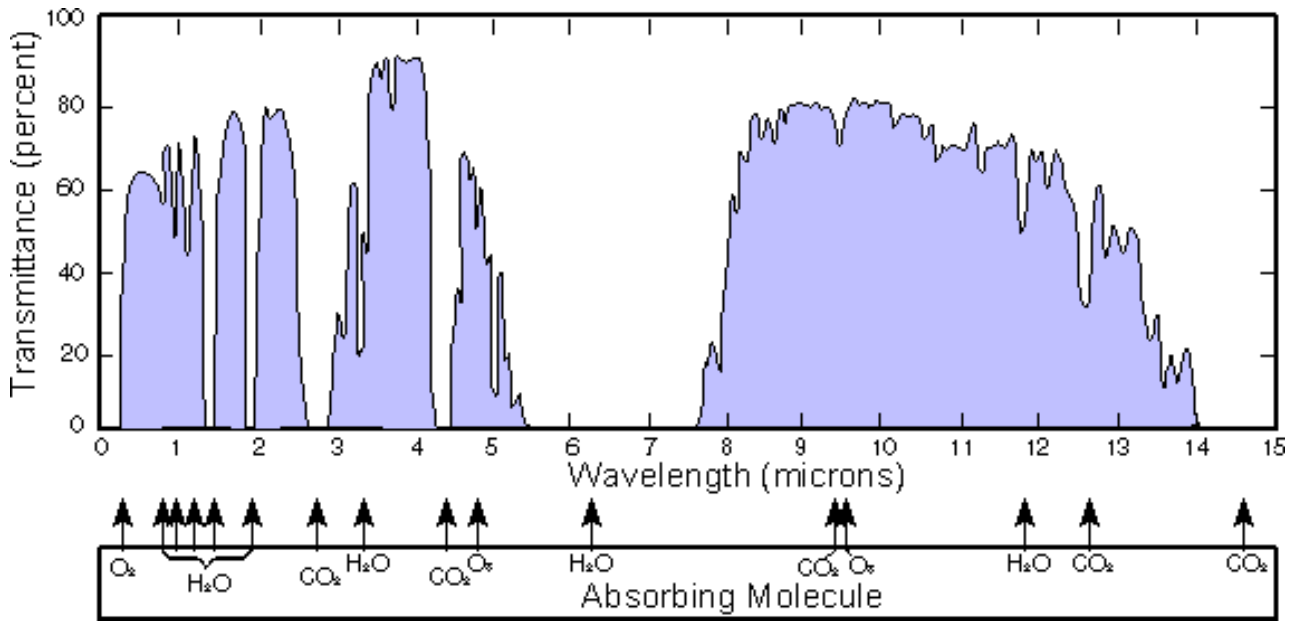

Fig. 5. (Color online) Atmospheric transmittance of light with wavelengths from 1–15 $\mu$m (Atm, 2006).

and an approximately 10 m long twin-engine propeller plane is barely detectable at a range of 7 km.

There are two cameras in the Dalek pointing towards the zenith. One camera is an FLIR Boson 640 infrared camera with a 95° FOV. The other camera is a visible/NIR ZWO ASI462MC CMOS color imager with $1936 \times 1096$ resolution and a 150° FOV (see Table 2 for full ZWO ASI462MC specifications).

#### 3.1.1. *Dome Environmental Control*

The exterior of the Dalek dome will be subject to a wide range of wind, temperature, humidity and illumination conditions which will affect the performance of the hardware within the dome. The structure of the Dalek is designed to isolate the cameras from stray external forces like wind, rain, animals, or fingers. A 42 mm stable platform provides a rigid support to the adjustable camera mounts. A finite element analysis of the structure has shown a maximum dome deformation of less than 1 mm of radial deformations when submitted to a 200 mph wind. A hurricane of category 5 produces wind speeds of up to 157 mph.

To protect the IR sensors from direct sunlight, the Dalek has 8 deployable shades mounted on servos, controlled by an Arduino micro-controller. These shades are depicted in blue in Fig. 4(a).

The Dalek has two, 72 cubic inch, desiccant containers to keep the enclosure dry. The desiccant is reusable and needs to be baked after changing color to regain its moisture-absorbing capabilities.

As the Dalek will be placed outdoors and will often be in direct sunlight, excess heat must be evacuated from the dome in order to protect the hardware inside. For example, the maximum operating temperature of the FLIR Boson 640 is 80°C. We have therefore placed a fan in the dome enclosure in order to shunt hot air out of the dome. The analysis provided in Appendix A demonstrates that this fan will be sufficient to cool the Dalek's hardware in most outdoor environments.

Table 2. ZWO ASI462MC camera specifications.

| Sensor | Sony IMX462 CMOS |
|---|---|
| Field of View | 150° |
| Resolution | 2.1 Mega Pixels<br>$1936 \times 1096$ pixels |
| Pixel size | $2.9\,\mu\text{m} \times 2.9\,\mu\text{m}$ |
| Max. Frame rate | 136 fps |
| Exposure time | From 32 $\mu$s to 2000 s |
| Wavelength range | 400 to 1000 nm |
| Working temperature | −5°C to 50°C |

### 3.2. *Alcor Optical All-Sky Camera*

The purpose of the Alcor camera is to provide a high-resolution wide-field view of the day or night sky in the optical range. The Alcor is an OMEA 9C Color camera with a $180° \times 180°$ FOV in the visible spectrum (350–750 nm). Like the Dalek, the Alcor is designed to be deployed outdoors and is robust to extreme weather. The camera's operating temperature ranges from −35°C to +45°C. The camera is protected by an acrylic dome which keeps out water, dust and insects; O-rings are used to ensure the dome is water-tight. The unit is equipped with temperature and humidity sensors and can automatically perform a dome defrost when necessary.

The full specifications of the Alcor camera are given in Table 3.

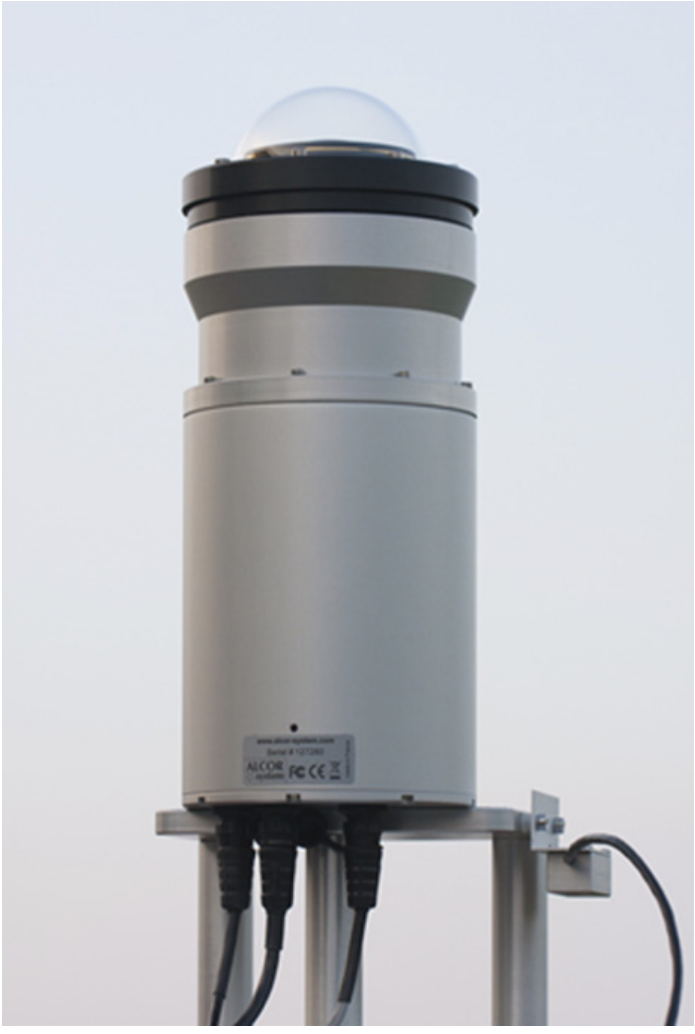

Fig. 6. Alcor all-sky camera.

Table 3. Alcor camera specifications.

| | |
|---|---|
| Field of View | $180^\circ \times 180^\circ$ |
| Resolution | $9575 \times 6380$ pixels |
| Pixel size | $3.8\,\mu\text{m} \times 3.8\,\mu\text{m}$ |
| Max. Frame rate | 1.5 fps USB 2.0<br>3.5 fps USB 3.0 |
| Gain | Variable, logarithmic mode from 30× to 10000× |
| Exposure time | From $32\,\mu$s to 1 h exposure all-sky mode, typically 20 to 75 s |
| Color? | Yes |
| Wavelength range | 350 to 750 nm (IR cut filter) |

### 3.3. *NPACKMAN Optical Wavelength Camera*

The New PArticle Counter k-index Magnetic ANomaly (NPACKMAN) instrument is a novel device designed specifically for the Galileo Project to provide data on environmental conditions and space weather near the location of deployment. This new unit is based on its predecessor, the open source PACKMAN designed for space weather detection (Mathanlal *et al.*, 2021).

The instrument contains an all-sky camera which captures images in the optical range. The specifications for this camera are given in Table 4.

### 3.4. *Data Transmission*

Each camera described above is physically linked (e.g. via USB cable) to an Nvidia Jetson (an "Edge" computer as described in Cloete (2023)). The Jetson will execute code which is responsible for performing initial object detection on each incoming camera image as described in Cloete (2023). If an object or objects are detected in a particular image, the Jeston will compose a message containing the centroid of the detection in image coordinates; the camera's id; the timestamp of the image; and a universally unique identifier (UUID) tagging the detected object. The composed message will be sent over a secure 0MQ (Zer, 2022) channel to a central processor; 0MQ is a popular open-source message-sending code library. In the first phase of the Project, this central processor will be the "Thinkmate" described in Cloete (2023). The object localization and kinematics analysis described in Sec. 4.5 will take place on this central processor. The purpose of the UUID is so that we can link the same detected object across multiple camera images taken by the same camera. This UUID will enable us to infer the velocity of localized objects, as described in Sec. 4.5.

Table 4. Characteristics of the NPACKMAN's optical camera.

| Sensor type | Sony Exmor RS CMOS IMX477 |
|---|---|
| Field of View | $170^\circ \times 170^\circ$ |
| Resolution | 12.3 Mega Pixels<br>$4056 \times 3040$ pixels |
| Pixel size | $1.55\,\mu\text{m} \times 1.55\,\mu\text{m}$ |
| Maximum Frame rate | 15 fps @$4056 \times 3040$ |
| Color? | Yes |

The Jetson hosts a Gstreamer (Gst, 2022) pipeline which packages image feeds from each connected camera into 30 s segments and securely transports these segments to the central processor for storage and asynchronous analysis. For the initial phase of this project, all image data will be transported to the central server regardless of whether a detection occurred. This decision allows us to use this image data to hone our detection and calibration algorithms for future phases. Once the central processor has acknowledged storage of a segment, the Jetson will delete the local segment file to recover storage space.

One Jetson will often receive data from more than one camera to minimize computing cost. For example, four Flir Boson cameras on the Dalek are routed to a single Jetson.

## 4. Software-Implemented Camera Calibration and Object Localization

### 4.1. *Intrinsic Visual, NIR, and IR Camera Calibration*

Intrinsic calibration of visual cameras is a well-solved problem. In this work, we use the chessboard calibration method provided by the OpenCV software library (Ope, 2022a). Each camera will be calibrated independently. The resulting intrinsic parameters (focal length, optical image center and distortion coefficients) will be stored in a central repository for fetching at runtime (i.e. during object localization). As we assume intrinsic parameters do not change over time (except when e.g. focal length changes as mentioned above), each camera's intrinsic parameters will be cached in the processor used by that camera to perform object detection.

The same chessboard calibration method can be adapted to determine the intrinsic parameters of IR and NIR cameras as described in Smisek *et al.* (2013).

We will take care to use OpenCV's calibration function for cameras using fish-eye lenses (Ope, 2022b) where appropriate. We define a fish-eye lens as in Kannala and Brandt (2006): a lens with a wide (about 180 degree) FOV whose projection model is non-perspective (i.e. stereographic, equidistance, equisolid angle, or orthogonal). The Alcor camera (see Sec. 3.2) for example has a fish-eye lens.

While we have stated that a camera's intrinsic parameters will not change over time, we cannot be sure this is the case. For example, temperature changes could deform our lenses sufficiently to warrant the re-computation of lens distortion coefficients. Thus, we will recompute each visual camera's intrinsic parameters at regular intervals in the field to confirm this assumption. If — given how a camera is sited in the field — it proves difficult to manually position the above-mentioned chessboard pattern close enough to the camera to compute acceptable intrinsic parameters, we will attach the chessboard pattern to a drone and fly the drone near the camera.

### 4.2. *Extrinsic Visual, NIR and IR Camera Calibration*

As described above, extrinsic camera calibration is the process of inferring the rotation matrix and translation vector which transforms a camera's coordinate frame to world coordinates.

We will use the Global Positioning System (GPS) to compute the translation vector of each camera, expressing the translation vector in the coordinate system employed by GPS: the World Geodetic System 1984 (WGS 84) (Wiley, 2009). GPS devices on smartphones are typically accurate to within approximately 5 m (GPS, 2022) but this can vary with terrain as well as proximity to buildings and trees. We show in Sec. 5 that this level of accuracy in our translation vector estimates often (though not always) leads to an acceptable level ($<100$ m) of localization error.

The determination of a camera's orientation with respect to the world coordinate system is less straightforward than the computation of its translation vector. Our cameras may be sited in locations which are hard for humans to reach. Also, a camera's orientation may drift over time as a camera is blown or jostled. We demonstrate in Sec. 5 that a relatively small inaccuracy in a camera's orientation estimate can lead to a large localization error. This source of triangulation error creates the requirement for low-cost, low-effort options to continuously verify the cameras' extrinsic calibration matrices, and adjust them if necessary. We describe three such techniques below.

#### 4.2.1. *Extrinsic calibration with ADS-B-equipped airplanes*

The US Federal Aviation Administration requires that most airplanes flying in US airspace be equipped with Automatic Dependent Surveillance — Broadcast (ADS-B) systems (FAA, 2022). With ADS-B, airplanes transmit their GPS-derived positions (latitude, longitude and altitude above mean sea level) in near real time. ADS-B receivers are readily available for consumer purchase.

Using data from an ADS-B receiver, we will compute a camera's extrinsic parameters using the known locations in the sky of airplanes imaged by that camera. We will use GPS to determine the position of a camera's origin with high accuracy. The question remains as to how to compute the rotational aspect of the extrinsic parameters for each camera.

The camera-to-be-calibrated will capture $N$ images of the airplane, associating with each image the airplane's ADS-B-provided 3D coordinates at the time of image capture. The ADS-B data will typically not be synchronized with image capture, so we will interpolate ADS-B data as necessary. Using these data, we will form two arrays of look-at unit vectors:

- $W$: the array of unit look-at vectors from the camera's known origin in world (GPS) coordinates to the airplane's position in those same world coordinates (as given by ADS-B) and;
- $C$: the array of unit look-at vectors from the camera to the airplane in camera coordinates. We will compute these camera look-at vectors by manually identifying the image point corresponding to the airplane's GPS receiver antenna in each image and then using the camera's intrinsic parameters to convert these image coordinates to a camera look-at vector in camera coordinates. If a particular airplane's GPS receiver location is not known, then we will chose an image point close to the centroid of the airplane. We will primarily use images of far-away, ill-resolved aircraft so the uncertainty stemming from GPS receiver location is minimized.

The $i$th vector in $W$ is denoted $w_i$; the $j$th vector in $C$ is denoted $c_j$. The rotation matrix $R$ we seek is that which maps $w_i$ to $c_i$ for all $i$, with minimal error. This problem of determining $R$ from vector observations is known as Wahba's problem (Wahba, 1965). As in Markley (1988), we estimate $R$ using singular value composition as follows:

(1) Compute the matrix $B = \sum_{i=1}^{N} w_i c_i^T$.
(2) Find the singular value decomposition of $B$: $B = USV^T$.
(3) Compute $M = \begin{pmatrix} 1 & 0 & 0 \\ 0 & 1 & 0 \\ 0 & 0 & \det(U)\det(V) \end{pmatrix}$
(4) The error-minimizing rotation matrix is $R = UMV^T$.

ADS-B-based extrinsic camera calibration appears to be a novel extrinsic calibration technique. As optical, NIR and IR cameras are all capable of imaging airplanes, the method described in this section can be used to extrinsically calibrate all cameras described in Sec. 3.

#### 4.2.2. *Extrinsic calibration with a GPS-equipped UAV*

This technique for computing a camera's extrinsic rotational parameters requires an unmanned aerial vehicle (UAV) equipped with a GPS receiver. The procedure we will use is inspired by the one outlined in Commun *et al.* (2021). The UAV will be flown manually within viewing range of a camera to be extrinsically calibrated. We will use the procedure described in Sec. 4.2.1 to infer a camera's translation vector and rotation matrix (of course substituting UAV for airplane). The position of the GPS receiver on the UAV will be manually identified in each image. We will add hardware to the drone so that it is visible to our IR and NIR cameras; we can use this method to to extrinsically calibrate all cameras described in Sec. 3.

#### 4.2.3. *Extrinsic calibration with celestial sky-marks*

Klaus *et al.* (2004) describe a method to perform both intrinsic and extrinsic camera calibration using a single image of the night sky taken by the camera to be calibrated. Given a night-sky image, the technique first isolates the centroids of bright stars in the image. (Note that in order to capture enough starlight, the commercial-grade cameras used in the study required 10–30 s exposure time.) The star associated with each image centroid is then identified. We will use software like AstroMB (Ast, 2022) to perform this star identification given the night-sky image as well as the longitude, latitude and time at which the image was taken. Once we have identified these stars, we can easily query their locations in the celestial sphere from publicly available star catalogues. Stellar locations in the celestial sphere are typically expressed using the equatorial coordinate system (ECS) (McClain and Vallado, 2001). Once we have the correspondences between 3D stellar locations and 2D image centroid positions, we can use the procedure outlined in Sec. 4.2.1 to compute the rotation matrix $R$ of each camera. Note that this rotation matrix will be defined with respect to the equatorial coordinate system.

Levit *et al.* (2008) also make use of this form of extrinsic calibration, though the authors do not describe their method in detail.

#### 4.2.4. *Combining extrinsic calibration estimates*

We will use all three techniques to maintain accurate extrinsic camera calibrations. As the procedures described in Secs. 4.2.1 and 4.2.2 involve manual input, we will execute the calibration procedures described therein on a weekly basis for each camera (and more frequently if possible given resource constraints). The technique described in Sec. 4.2.3 is fully automatable so we will execute it every night for each relevant camera (unless cloud cover obscures the night sky).

The procedures described in Secs. 4.2.1 and 4.2.2 employ the WGS 84 coordinate system whereas the star-based calibration technique outlined in Sec. 4.2.3 uses the equatorial coordinate system; we will have to translate between the two. We can use the ogr2ogr (ogr, 2022) program to perform this conversion. We will standardize to the WGS 84 coordinate system.

The question remains as to how we will combine the rotation matrices provided by the three calibration methods described above for a single camera. We will do this as follows: every time a new rotation matrix is generated, we will take that new matrix and the two most recent matrices generated by the other two methods and combine them using the rotation matrix "averaging" method described in Bhardwaj *et al.* (2018).

### 4.3. *Infrared-Specific Camera Calibration*

In addition to the above-described intrinsic and extrinsic calibration routines which all of the Project's cameras must undergo, our FLIR Boson infrared cameras (Sec. 3.1) require additional calibration steps to remove image non-uniformities (INUs). These calibration procedures are described below.

#### 4.3.1. *Calibration for INU removal*

INUs are "defined as a group of pixels which are prone to varying slightly from their local neighborhood under certain imaging conditions" (Fli, 2018). INUs can be caused by such artifacts as dirt or droplets on the Boson's outer protective window (which comes with the camera), or small defects on its inner protective window. A guide provided by the Boson manufacturer (Fli, 2018) describes two methods to mitigate the impact of INUs: lens gain correction and supplemental flat-field correction (SFFC).

Lens gain correction is computed by imaging two scenes at different, uniform temperatures (Fli, 2017). The manufacturers of the camera provide software to calculate and store the lens gain correction given these two images. While lens-gain correction is performed before each camera is shipped, the manufacturer recommends that Boson owners repeat this correction after mounting the Boson on-site, or if the camera's focal length changes significantly.

Supplemental flat-field correction is used to account for the heat generated internally by the camera. As the camera's heating characteristics could change after the camera is installed (e.g. if camera heat sinks are installed), the manufacturer recommends that SFFC be carried out post-installation. SFFC requires a blackbody source fully occupying the camera's FOV. This procedure need to be only carried out once after installation (Fli, 2017).

Lens gain should be carried out before supplemental flat-field correction.

### 4.4. *Image Data Synchronization*

As described in Sec. 2.1, each camera must closely synchronize the capture of images. We show in Sec. 5 that mis-synchronization of image capture is a non-trivial source of object localization error.

Suppose there are two cameras: A and B. Camera A takes a series of images $a_1, a_2, a_3, \ldots$ and camera B captures a series of images $b_1, b_2, b_3, \ldots$. In order to use these images for object localization, we need to know the image $a_i$ which was captured closest in time to any image $b_j$. We shall call these images $a_i$ and $b_j$ "nearly simultaneous" in the remainder of this work.

We will extend this procedure to all pairs of cameras whose images we will use for object localization.

We will use the Network Time Protocol (NTP) to synchronize as closely as possible the clocks on all computers to which our cameras are connected. Mills (1989) reports that NTP as originally implemented typically yields an accuracy on the order of tens of milliseconds with respect to the time reference source.

In addition to synchronizing computer clocks, we also need to match as closely as possible the times at which different cameras capture individual images. We will focus in the initial phase of this project on synchronizing image capture for same-brand cameras (e.g. for all Alcor cameras). As described in proceeding sections, these same-brand cameras will be connected to different computers in the field and will be separated by distances of several kilometers. The frame rates for these same-brand cameras will be set identically for each camera. Our method for synchronizing image capture for these multiple, non-co-located, same-brand cameras will be primitive: we will simply continue to restart the image capture programs for each camera until those cameras differ in image capture times by not more than a configurable time difference $d_{\text{time}}$ (in seconds). This target time difference will be set depending on how much localization error we are willing to tolerate from non-synchronized image capture. We will estimate the function of synchronization error to localization error for a given camera brand with experiments like those described in Sec. 5.

Finally, for same-brand cameras, we will decide that an image $a_i$ from camera A and an image $b_j$ from camera B are nearly simultaneous if the absolute difference in timestamps between $a_i$ and $b_j$ is less than $d_{\text{time}}$.

### 4.5. *Object Localization with Multiple Synchronized Images*

We assume all cameras are calibrated as described in Sec. 2.1. We define a camera suite as a set of proximate Dalek, Alcor, and NPACKMAN cameras. For the purposes of localization and tracking, at least two camera suites will be deployed at least some distance apart; we explore suitable camera separation in Sec. 5. Not every camera suite will necessarily house all of the cameras described above. Each camera in each suite will independently use the object detection algorithm described in Cloete (2023) to detect the bounding boxes and centroids of objects of interest in the stream of images captured by these cameras. Each detection in each image will be labeled as described in Sec. 4.4 so that we can associate images taken from multiple cameras in a single localization computation.

For each detection in each image, we will send a 0MQ message to our central processor on a secure messaging channel called "VisualDetection FromImage" consisting of the centroid of the detection in image coordinates; the camera's id; the timestamp of the image; and a universally unique identifier (UUID) tagging the detected object as described in Sec. 3.4.

When the central processor receives a detection message on the channel "VisualDetection FromImage", it will read the message and

(1) undistort the detected centroid using the camera's computed intrinsic parameters as described in Sec. 2.1;
(2) look up the camera's camera suite id in the central processor's database; and
(3) send a new 0MQ message consisting of the contents of the original message along with the undistorted image point in image coordinates and the camera suite id on a secure 0MQ messaging channel called "AugmentedVisual Detection".

The central processor will subscribe to the "AugmentedVisualDetection" channel. After receiving an initial message on the "AugmentedVisual Detection" channel, the central processor will wait for a short, configurable period of time for other messages from other cameras (both cameras from the suite associated with the first camera and from all other camera suites) whose detections occur nearly simultaneously with the detection encoded in that initial message. We shall call the set of messages gathered in this way the "detection event set." Each camera can contribute at most one image to a detection event set.

All "AugmentedVisualDetection" messages will be written to a secure persistent store accessible by our central processor in the initial phase of this project for debugging purposes.

Given a detection event set, there are two problems we must solve in order to localize the detected objects. We must first decide which 2D image detections relate to the same 3D object; this is the correspondence problem. Once we have assigned an object to each detection, we must infer that object's 3D location given the set of detections (and in particular the 2D image coordinates of those detections along with the intrinsic and extrinsic parameters of the detectors); this is the triangulation or localization problem. We describe our solution to the triangulation problem first, and follow with a description of our solution to the correspondence problem.

#### 4.5.1. *Object triangulation*

Generally, we may have more than two detections for the same object (i.e. when we are employing more than two cameras for triangulation). In that case, Hartley & Zisserman (2004) recommend (in Sec. 12.2) localizing that object using the homogeneous linear triangulation method (LTM-H). The authors state that LTM-H "often provides acceptable results [and] has the virtue that it generalizes easily to triangulation when more than two views of the [3D] point are available," unlike other methods.

LTM-H works as follows: Suppose $X$ is the (unknown) 3D position of an object in world coordinates and $X_i = (x_i, y_i)$ is the (known) 2D image point of that object as captured by a camera with projection matrix $P$ (as described in Sec. 2.1). It holds that $X_i = PX$. Since the cross-product of $X_i$ and $PX$ should be equal to 0 for the correct value of $X$, we can form three equations which are linear in $X$

$$x_i(p^{3T}X) - (p^{1T}X) = 0, \tag{9}$$

$$y_i(p^{3T}X) - (p^{2T}X) = 0, \tag{10}$$

$$x_i(p^{2T}X) - y_i(p^{1T}X) = 0. \tag{11}$$

We can re-write Eqs. (9) and (10)[a] as

$$AX = 0, \tag{12}$$

where $A$ is the matrix

$$\begin{pmatrix} x_i p^{3T} - p^{1T} \\ y_i p^{3T} - p^{2T} \end{pmatrix} \tag{13}$$

and $p^{nT}$ denotes the $n$th row of $P$. In matrix $A$, we can "stack" further image points for the same object as seen by other cameras. For example, if the same object is imaged at point $(x'_i, y'_i)$ in a camera with projection matrix $P'$ and imaged at point $(x''_i, y''_i)$ in a camera with projection matrix $P''$, then we can write $A$ as

$$\begin{pmatrix} x_i p^{3T} - p^{1T} \\ y_i p^{3T} - p^{2T} \\ x'_i p'^{3T} - p'^{1T} \\ y'_i p'^{3T} - p'^{2T} \\ x''_i p''^{3T} - p''^{1T} \\ y''_i p''^{3T} - p''^{2T} \end{pmatrix}. \quad (14)$$

[a]Note that Hartley & Zisserman (2004) suggest using only Eqs. (9) and (10) here and not Eq. (11). We have followed suit.

As described in Hartley & Zisserman (2004), Eq. (12) will be solved by first computing the singular value decomposition of $A$: $U\Sigma V^T$. The solution $X$ (expressed as a homogeneous vector) will be the fourth column of $V$.

We will additionally introduce the reprojection error (Hartley & Zisserman, 2004) for the solution $X$ as follows: For each camera used in the solution, compute $PX = \hat{x}$; $\hat{x}$ is the reprojection of $X$ onto the camera with projection matrix $P$. The reprojection error for this image is the Euclidean distance between $\hat{x}$ and $(x_i, y_i)$. The total reprojection error for $X$ is the average of the reprojection errors over all images used in the estimation of $X$.

#### 4.5.2. *Object correspondence in multiple images*

Given a detection event set, we will use the Constraint Satisfaction Optimization Problem (CSOP) (Tsang, 1993) framework to determine the optimal image correspondences for that detection event set. CSOP is a generic framework which assigns values to a defined set of variables while satisfying a set of defined constraints. The solution to a CSOP is the variable assignment which satisfies those constraints while maximizing (or minimizing) a given optimization function.

Our variables in this case are the set of detections within the detection event set. Let $C_iD_j$ be the variable denoting the $j$th object detected in the image from camera $i$. Each such variable is assigned an object identifier $O_k$, where $O_k$ is a potential 3D object to be localized. In the most extreme case, each detection in the detection event set will correspond to a distinct object so the total number of possible object identifiers is the total number of detections across all camera images (i.e. $1 \leq k \leq \sum C_iD_j$).[b]

[b]In the extreme case where each detection corresponds to a distinct object we will not be able to localize any of those objects, but this case is possible in the real world. For example, consider the case where each camera captures a nearby insect, but these insects are too small and distant for other cameras to image.

For our problem, the relevant CSOP constraints are as follows:

(1) An object $O_k$ can only be assigned at most once per camera image $C_i$. In other words, a camera cannot detect the same object more than once in the same image.
(2) An object $O_k$ can only be assigned to a pair of detections $C_iD_j$ and $C_mD_n$ $(i \neq m)$ if that assignment provides a pair of world look-at vectors whose closest point of intersection is (a) within viewing range of each camera $C_i$ and $C_m$ assuming a minimum object size of 30 m and, (b) not behind either camera. We chose this minimum object size as this is a typical length of small aircraft.
(3) At least one object in any valid solution must be imaged by 2 or more cameras.
(4) Any given object must be imaged by cameras from at least two distinct camera suites.[c]

Without loss of generality — to cut down the search space of our problem's solution set — we will assign the camera image with the most detections the first $D$ objects, where $D$ is the number of detections in this camera image.

Finally, our CSOP optimization algorithm is as follows: For each candidate solution (i.e. assignment of each variable $C_iD_j$ to an object $O_k$) which satisfies the above constraints, we find each object in the solution which has two or more assigned detections and form the matrix $A$ from the image points related to those detections along with the projection matrix of each camera whose image points we are using. We use LTM-H to solve for $X$ in Eq. (12). $X$ is then our candidate object location. We repeat this for all objects assigned in our current CSOP solution. We compute the reprojection error for each object in the current CSOP solution. For the purposes of optimization, the maximum reprojection error over all objects in the current CSOP solution is the error for that CSOP solution. If the error of a CSOP solution is less than a threshold $T_p$ pixels (where $T_p$ is a configurable parameter which will be determined empirically), then we consider the localization to be sound. Our accepted solution overall is that with the highest number of sound object localizations. If multiple candidate solutions have the maximum number of sound object

[c]Note that this does not rule out solutions in which multiple cameras from the same suite also detect this object.

localizations then we accept the solution with the smallest average reprojection error.

The detection information (detection UUIDs, detection timestamps, image points and projection matrix for each camera) for each successful object localization will be stored in our central processor's persistent store. This information will be used to compute object velocity and acceleration as described in Sec. 4.6.

### 4.6. *Velocity and Acceleration Computation*

Given two successive 3D localizations $X_1$ and $X_2$ of the same object at two distinct times $t_1$ and $t_2$ ($t_2 > t_1$), our central processor will compute the average velocity vector $V$ of that object over the time period in question:

$$V = \frac{X_2 - X_1}{t_2 - t_1}. \tag{15}$$

In order to use Eq. (15), we must answer the following questions:

- What is the time for which $X_1$ was computed and what was the time for which $X_2$ was computed?
- How do we know that $X_1$ and $X_2$ are locations of the same object?

The time for which $X_1$ (or $X_2$) was computed will be taken as the average of the times at which each image used to compute $X_1$ (or $X_2$) was captured.

The computation of $X_1$ as described in Sec. 4.5 will involve a set of object detections in images. Each detection will be tagged with a UUID. The set of detection UUIDs will provide an object signature to identify that object over time. As described above, the detection information used for each object localization will be stored in our central processor's persistent store. To determine which localization record in our persistent store is the next localization for that object (i.e. which stored record provides $X_2$), we will query the store for the localization record subsequent to $X_1$ which overlaps by at least one detection UUID with $X_1$'s set of detection UUIDs.

Each computed object velocity will be stored in our central processor's persistent store. Once we have three velocity calculations for a given object, we will compute the average acceleration of that object over the corresponding time period.

## 5. Experiments, Results and Discussion

### 5.1. *Localization Experiments*

Here, we describe simulation experiments to aid in determining the localization error we can expect given our camera suites, distance between camera suites, calibration methods, and triangulation algorithm described in the previous sections. This set of experiments will assume that the correspondence problem is solved.

A basic question to answer in setting up these experiments is: how far apart should two camera suites be located in order to achieve a given target localization accuracy? We shall call this distance the baseline for a camera pair. Our investigation of this question employs the following equation from Gallup *et al.* (2008)

$$\epsilon_z \approx \frac{z^2}{bf}\epsilon_d, \tag{16}$$

where

- $z$ is the distance of the localized position from the baseline;
- $\epsilon_z$ is the error in the distance estimate $z$;
- $b$ is the camera baseline length (i.e. the distance between the pair of cameras);
- $f$ is the camera focal length (assumed to be equal for each camera in the stereo pair); and
- $\epsilon_d$ is the correspondence error (assumed to be equally divided between the two images). The correspondence error can be understood as the difference (in pixel units) between where an image of an object ought to be given no imaging error versus where the image of the object actually is. For example, if one camera in a stereo pair takes its image of a moving object after the other camera, this will induce a correspondence error. (In this case, the distance error induced by this situation was explored in Eq. (8).)

The camera setup for which Eq. (16) was derived is different than the setup we will employ in this Project. For example, the focal lengths for two cameras providing corresponding image points for localization may in reality be different. Also, more than two cameras may be employed when estimating object location.

Equation (16) describes distance-from-baseline error while in this Project we are interested in localization error. Nonetheless, despite these differences, we will use Eq. (16) to provide an initial rough estimate of the camera baselines we should use in our experiments.

Equation (16) can easily be converted to

$$b \approx \frac{z^2}{\epsilon_z f}\epsilon_d \tag{17}$$

which will allow us to estimate–for a given pair of cameras with identical focal lengths, a given maximum desired object-to-baseline distance $z = z_{\max}$ and a given correspondence error $\epsilon_d$ — what baseline distance between cameras is required?

We will set $z_{\max}$ equal to 15 km since — as discussed in Sec. 3.1 — our FLIR Boson cameras are capable of resolving mid-sized objects (20 m long or greater) at this distance. We will explore various settings of $\epsilon_z$: 1 m, 10 m and 100 m. The distribution of $\epsilon_d$ is unknown and depends on several factors:

- The accuracy of the YOLO algorithm (Cloete, 2023) in producing image centroids for the same object from different cameras.
- The factors which are used to compute each camera's projection matrix: the camera's intrinsic and extrinsic parameters.
- The accuracy with which we "undistort" images.
- For objects which are in motion with respect to our camera pairs, the time difference between image capture between the two cameras.

In producing our rough estimates for $b$, we will explore values of $\epsilon_d$ of 0.1 pixels, 1 pixel, and 10 pixels. The focal length $f$ depends on the two cameras used to estimate distance-from-baseline. For rectilinear lenses $f$ is related to the FOV and the image sensor diagonal size $d$ by

$$f = \frac{d}{2\tan\left(\frac{\text{FOV}}{2}\right)} \tag{18}$$

as given in Bettonvil (2005). For fish-eye lenses with equisolid projections, the following relationship holds:

$$f = \frac{d}{2\sin\left(\frac{\text{FOV}}{2}\right)} \tag{19}$$

as given in Bettonvil (2005). We will assume the two cameras are identical and therefore have the same focal length when estimating $b$. The focal length of our Project's cameras is given in Table 5.

Given focal length data in Table 5 and Eq. (17), we have computed the required camera baseline lengths in Table 6 for various values of $\epsilon_z$ and $\epsilon_d$. For example, for a desired distance error ($\epsilon_z$) of 10 m and an assumed correspondence disparity ($\epsilon_d$) of 0.1 pixels, a pair of Alcor cameras should be separated by about 300 m.

Table 5. Camera focal lengths. Above, "FEE" denotes "Fisheye Equisolid" and "RL" denotes "Rectilinear."

| Camera | Lens | FOV (deg.) | $d$ ($\mu$m) | $f$ ($\mu$m) |
|---|---|---|---|---|
| Alcor | FEE | 180 | 43700 | 15400 |
| FLIR Boson 640 | RL | 50 | 9830 | 8700 |
| FLIR Boson 640 | RL | 95 | 9830 | 4900 |
| ZWO ASI462MC | FEE | 150 | 6460 | 2650 |
| NPACKMAN | FEE | 170 | 8860 | 1700 |

Now that we have realistic baseline distances to choose from, we will determine the localization error we can expect given the camera suites, baseline distance between camera suites, calibration methods, and triangulation algorithm described in the previous sections. In order to do so, we will use a baseline camera distance of 1500 m. Table 6 indicates that this is a reasonable setting for all cameras, given a target distance error of 10 m and a correspondence error of 0.1 pixels.

As was done in Rumpler *et al.* (2011), we will use Monte Carlo simulation to compute the expected localization error given the following input random variables:

- $C$: the location of each camera origin as computed by extrinsic calibration. Each extrinsic calibration procedure defined above uses GPS to provide $C$. We will assume that error in measuring $C$ is zero mean Gaussian distributed with standard deviation of 5 m (from Izet-Unsalan & Unsalan (2020)). The 3D direction of the error will be chosen from a uniform distribution of unit vectors pointing in all directions.
- $R$: the orientation of each camera coordinate frame with respect to the world coordinate frame as computed by extrinsic calibration. Klaus *et al.* (2004) report an (anecdotal) error in their method of about 3°. We will assume that the error in measuring $R$ is Gaussian distributed with a zero mean error and a standard deviation 3° about the $z$-axis.
- We will assume that the YOLO algorithm described in Cloete (2023) induces a centroid position error that is Gaussian distributed with a mean error of 0 pixels (standard deviation 2) in a direction within the image plane which is uniformly distributed.
- We will assume that triangulated objects are moving parallel to the baseline of the camera pair

Table 6. Stereo camera baseline distances (in meters) required for a variety of camera types, correspondence errors, and target distance errors. The baseline distances have been rounded up to the nearest hundred.

| Camera | $\epsilon_z$ (m) | $\epsilon_d$ (pixels) | Required baseline (m) |
|---|---|---|---|
| Alcor | 1 | 0.1 | 2500 |
| Boson 50 FOV | 1 | 0.1 | 13800 |
| Boson 95 FOV | 1 | 0.1 | 24600 |
| NPACKMAN | 1 | 0.1 | 14200 |
| ZWO | 1 | 0.1 | 11000 |
| Alcor | 1 | 1 | 24600 |
| Boson 50 FOV | 1 | 1 | 138000 |
| Boson 95 FOV | 1 | 1 | 245100 |
| NPACKMAN | 1 | 1 | 141200 |
| ZWO | 1 | 1 | 109500 |
| Alcor | 1 | 10 | 245900 |
| Boson 50 FOV | 1 | 10 | 1379400 |
| Boson 95 FOV | 1 | 10 | 2451000 |
| NPACKMAN | 1 | 10 | 1411800 |
| ZWO | 1 | 10 | 1094400 |
| Alcor | 10 | 0.1 | 300 |
| Boson 50 FOV | 10 | 0.1 | 1400 |
| Boson 95 FOV | 10 | 0.1 | 2500 |
| NPACKMAN | 10 | 0.1 | 1500 |
| ZWO | 10 | 0.1 | 1100 |
| Alcor | 10 | 1 | 2500 |
| Boson 50 FOV | 10 | 1 | 13800 |
| Boson 95 FOV | 10 | 1 | 24600 |
| NPACKMAN | 10 | 1 | 14200 |
| ZWO | 10 | 1 | 11000 |
| Alcor | 10 | 10 | 24600 |
| Boson 50 FOV | 10 | 10 | 138000 |
| Boson 95 FOV | 10 | 10 | 245100 |
| NPACKMAN | 10 | 10 | 141200 |
| ZWO | 10 | 10 | 109500 |
| Alcor | 100 | 0.1 | 100 |
| Boson 50 FOV | 100 | 0.1 | 200 |
| Boson 95 FOV | 100 | 0.1 | 300 |
| NPACKMAN | 100 | 0.1 | 200 |
| ZWO | 100 | 0.1 | 200 |
| Alcor | 100 | 1 | 300 |
| Boson 50 FOV | 100 | 1 | 1400 |
| Boson 95 FOV | 100 | 1 | 2500 |
| NPACKMAN | 100 | 1 | 1500 |
| ZWO | 100 | 1 | 1100 |
| Alcor | 100 | 10 | 2500 |
| Boson 50 FOV | 100 | 10 | 13800 |
| Boson 95 FOV | 100 | 10 | 24600 |
| NPACKMAN | 100 | 10 | 14200 |
| ZWO | 100 | 10 | 11000 |

and we will assume that these objects are moving with a speed which is "very fast", "fast", "slow" or "stationary." "Slow" models an aerial drone moving at 1 meter per second. "Fast" models an airliner moving at 250 meters per second. "Very fast" models an object moving at 1000 meters per second.

- Triangulated objects will be uniformly distributed along two 3D lines. One line–the "far" line — is at a distance of 15 km above the baseline between our simulated stereo cameras; it is oriented parallel to the baseline. The other line — the "near" line — is at a distance of 0.1 km above the baseline and is oriented identically to the "far" line. Each line is 10 km long and its center is positioned above the center point of the baseline. The reason for choosing this distribution of object locations is that objects that are distant from the cameras will incur a distance estimation error as predicted by Eq. (16); fast-moving objects that are relatively close to the cameras will incur a relatively large correspondence disparity since images from each camera will be captured at different times, and the object will have moved in the interim.

- The time between image capture by each camera in the stereo pair will be uniformly distributed between 0 s and $\max_t$ s, inclusive. The value $\max_t$ is equal to the inverse of the maximum frame rate divided by 2 (as this is the maximum that each camera could be "out-of-sync" with respect to its stereo-pair cousin if both are run at the maximum frame rate allow).

Note that we have chosen not to include errors in intrinsic camera parameter estimates in this simulation as we assume these estimates will be derived in the lab in controlled conditions and will be highly accurate.

We will simulate pairs of Alcor cameras as well as pairs of ZWO ASI462MC cameras as these "All Sky" cameras have the widest fields of view and are most likely to capture the same object when deployed as stereo pairs. In simulation, each camera will capture and process images at its maximum frame rate. In each simulation iteration, we will choose values for the input random variables according to their distributions. The settings of random variables in successive simulation iterations will be independent from one another. One object will be placed and triangulated in each simulation iteration, and the error in that triangulation will be measured.

Our first simulation simply runs the triangulation algorithm without including any errors stemming from calibration, object movement or YOLO object centroid identification. The only source of error is the localization algorithm itself. The histogram of localization error for the Alcor and ZWO ASI462MC cameras are given in Fig. 7. For both camera types, the localization error is quite low ($\ll 1$ m). The localization error due to our chosen localization algorithm cannot be altered without changing the algorithm; this source of error is therefore present in all subsequent experiments.

Our second simulation runs the triangulation algorithm with errors only with knowledge of the location $C$ of the first camera in the camera pair. The histogram of localization error for the Alcor and ZWO ASI462MC cameras is given in Fig. 8. The distributions in Figs. 8(a) and 8(b) are unimodal, with modes quite close to zero. That said, large

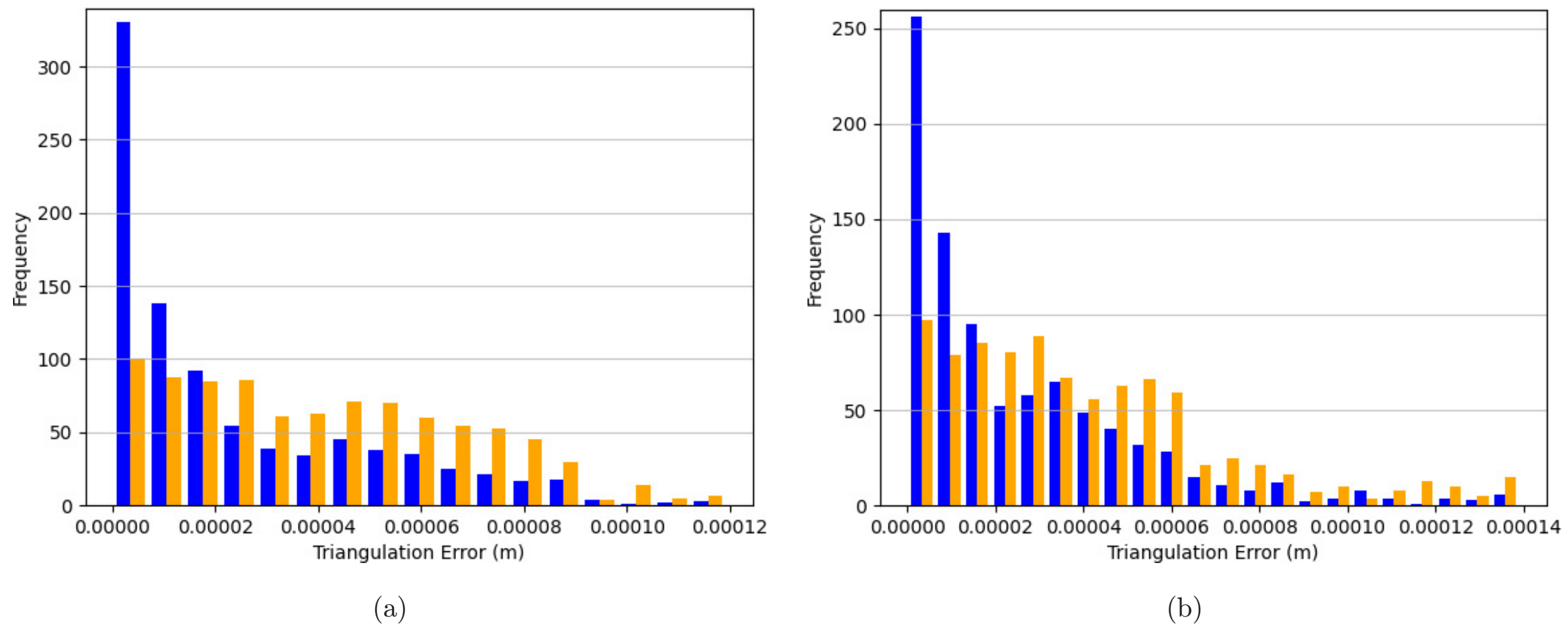

Fig. 7. (Color online) Histogram of localization errors for a pair of Alcor cameras (a) and ZWO cameras (b). The only errors introduced in the simulation were due to the triangulation algorithm itself. Blue bars represent localization errors for objects in the near field (100 m from the camera baseline). Orange bars represent localization errors for objects in the far field (15 km from the camera baseline). For each camera type, a total of 1792 simulation iterations were run, with 896 simulation iterations with far-field objects and 896 simulation iterations with near-field objects. Unless otherwise noted, the meanings of the histogram bar colors as well as the number of simulation runs remain the same for each subsequent experiment.

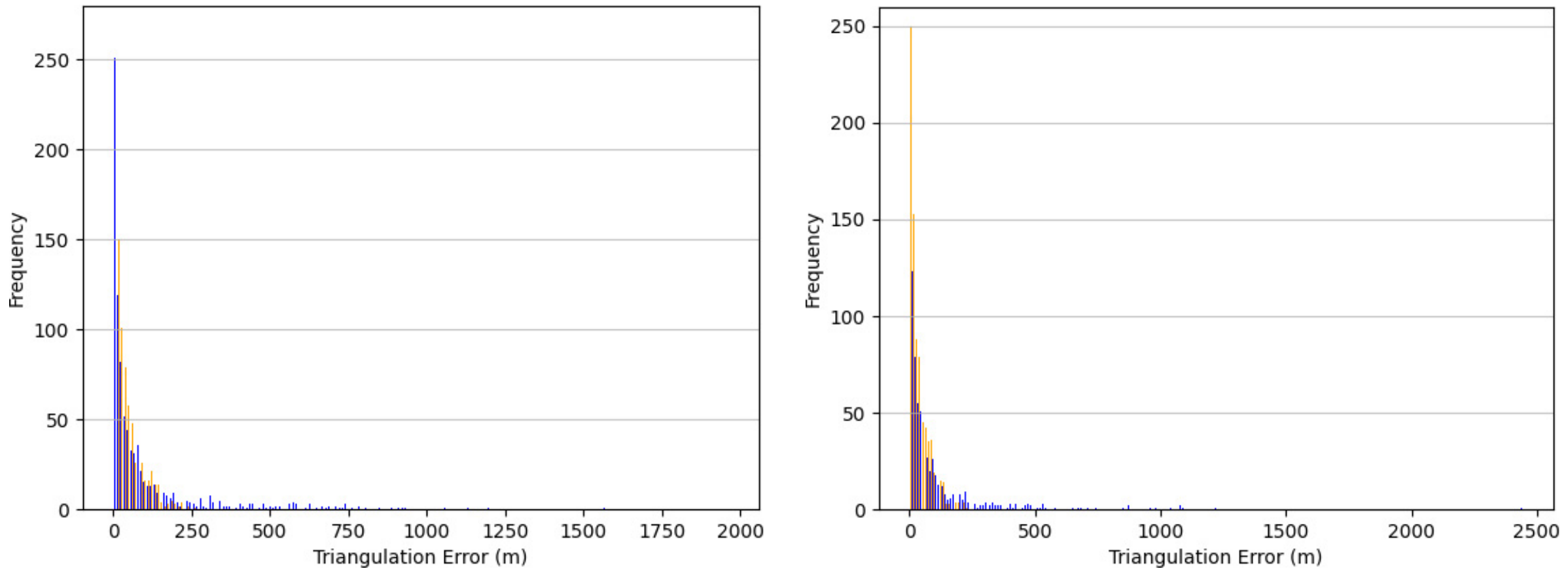

(a) (b)

Fig. 8. (Color online) Histogram of localization errors for a pair of Alcor cameras (a) and ZWO cameras (b). The errors introduced in the simulation were due to incorrect knowledge of the location of the first camera in the stereo pair (i.e. due to miscalibration), with an error distribution as described in the text. Note the difference in scale of the $x$-axis in the two graphs.

errors due to camera localization do occur. The largest errors occur with near-field objects.

Our third simulation runs the triangulation algorithm with errors only with knowledge of the orientation $R$ of the first camera in the camera pair. The histogram of localization error for the Alcor and ZWO ASI462MC cameras are given in Fig. 9. The distributions in Fig. 9(a) and 9(b) are unimodal, with modes that are quite large ($10^4$ m). Camera orientation uncertainty is by far the largest contributor of localization error found so far.

Our next simulation runs the triangulation algorithm with errors due to the fact that the object being localized is moving, and the cameras do not capture images at the same time. The histogram of localization error for the Alcor and ZWO ASI462MC cameras are given in Fig. 10. The distributions in Figs. 10(a) and 10(b) are unimodal, with modes quite close to zero. The Alcor camera has a much wider error distribution than does the ZWO camera; this is due to the fact that the Alcor's maximum frame rate is approximately 100 times smaller than the ZWO.

For this simulation run we also collected data showing localization error as a function of image capture time differences (Fig. 11). As discussed in Sec. 4.4 we can use plots such as this to determine the maximum acceptable difference in time between image capture by stereo camera pairs used to localize the same moving object. In the case shown in

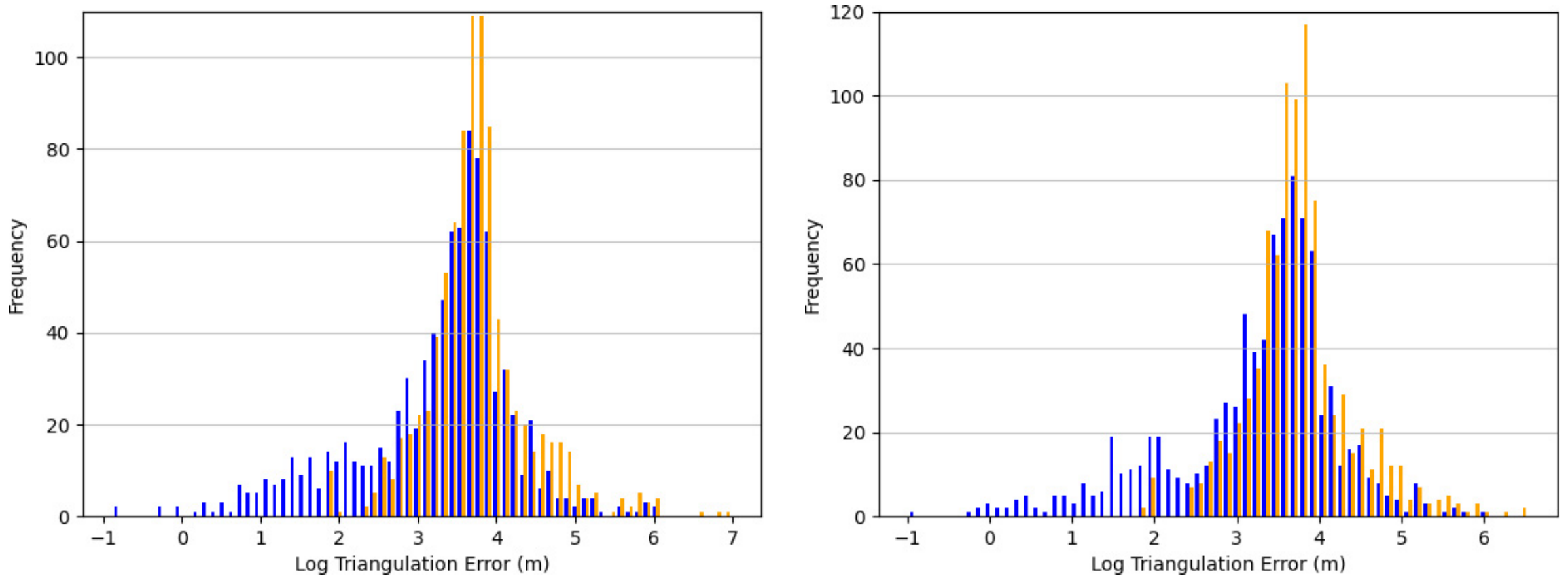

(a) (b)

Fig. 9. (Color online) Histogram of localization errors for a pair of Alcor cameras (a) and ZWO cameras (b). The errors introduced in the simulation were due to incorrect knowledge of the orientation of the first camera in the stereo pair (i.e. due to miscalibration), with an error distribution as described in the text. Note that the $x$-axis in the two graphs is presented in a log (base 10) scale.

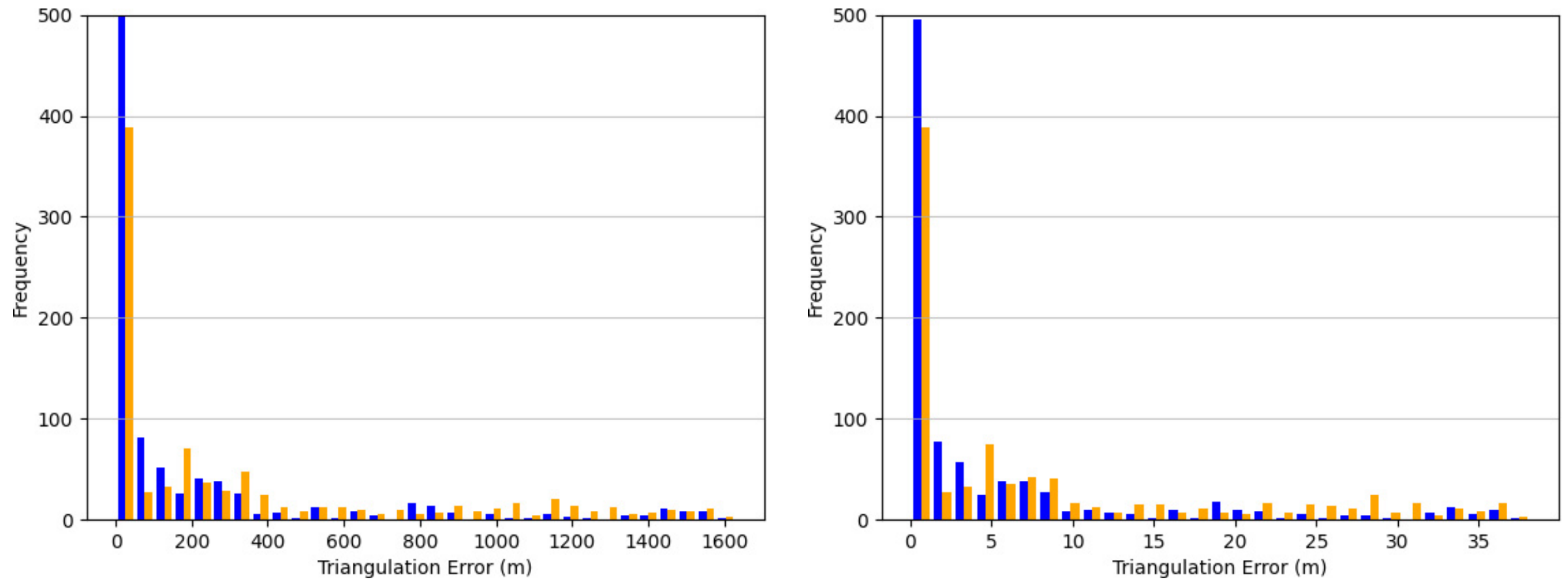

(a) (b)

Fig. 10. (Color online) Histogram of localization errors for a pair of Alcor cameras (a) and ZWO cameras (b). The errors introduced in the simulation were due to the fact that the localized object was moving and the second camera captured its image after the first camera, with a distribution of object speeds as described in the text. Note the difference in scale of the $x$-axis in the two graphs.

Fig. 11, image capture times must differ by not more than about $10^{-3}$ s in order to achieve localization accuracy of 10 m (assuming image capture time difference is the only source of such error).

Our next simulation runs the triangulation algorithm with errors introduced by our object detection algorithm (YOLO). The histogram of localization error for the Alcor and ZWO ASI462MC cameras is given in Fig. 12. The distributions in Figs. 12(a) and 12(b) are unimodal, with modes quite close to zero. The Alcor camera is less sensitive to detection errors than the ZWO because the Alcor has a higher resolution image sensor.

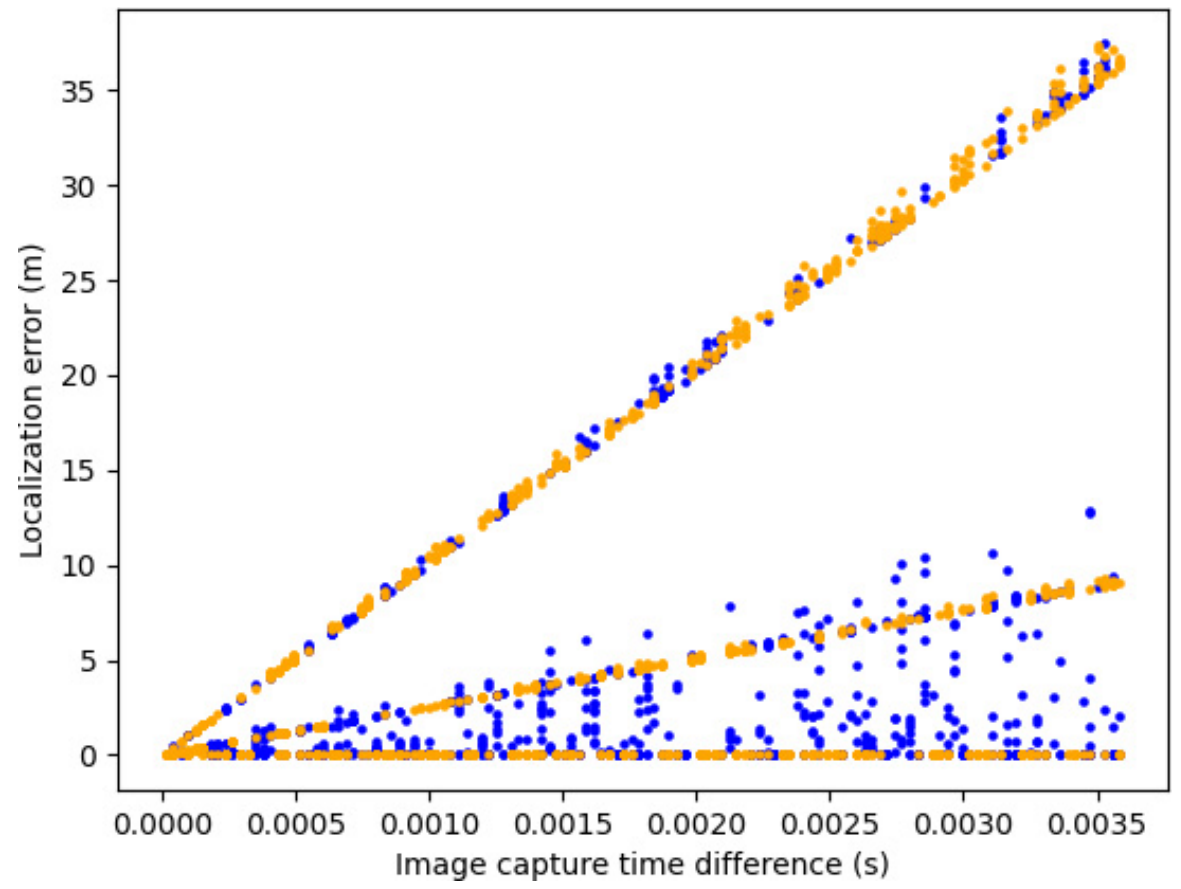

Fig. 11. (Color online) Scatter plot of localization errors (in meters) for a pair of ZWO cameras as a function of the time difference between image capture (in seconds). As usual, orange represents data for far-field objects and blue represents data for near-field objects.

We then simulated the triangulation algorithm with all of the sources of error described above simultaneously. The histogram of localization error for the Alcor and ZWO ASI462MC cameras are given in Fig. 13. The distributions in Figs. 13(a) and 13(b) qualitatively similar to the error distributions given in Fig. 9, indicating that errors in camera orientation are the main source of localization error.

### 5.2. *Correspondence Experiments*

In this set of experiments, we introduced multiple objects-to-be-tracked in the same simulated world, thus requiring the correspondence problem to be solved with our CSOP algorithm.

We simulated three cameras with origins at locations (0 m, −1000 m, 0 m), (0 m, 1000 m, 0 m), and (−1000 m, 0 m, 0 m). Each camera was oriented such that its optical axis pointed to the local zenith. We set $T_p$ to 1 pixel for all simulation runs.

In the first experiment, we inserted two objects in the simulated scene; the position of each object was assigned uniformly random values in the range $-2000\,\text{m} \leq x, y \leq 2000\,\text{m}$ and $200\,\text{m} \leq z \leq 2000\,\text{m}$. We ran 100 simulations with this setup. The image to object correspondences occurred with 100% accuracy in this case.

We then repeated this first experiment, but this time we perturbed the calibrated orientation of the third camera by 1° about its optical axis. This simulates an error due to e.g. the miscalibration of the third camera. In all 100 runs, detections from this third camera were successfully discarded as erroneous and ignored in the computation of object location.

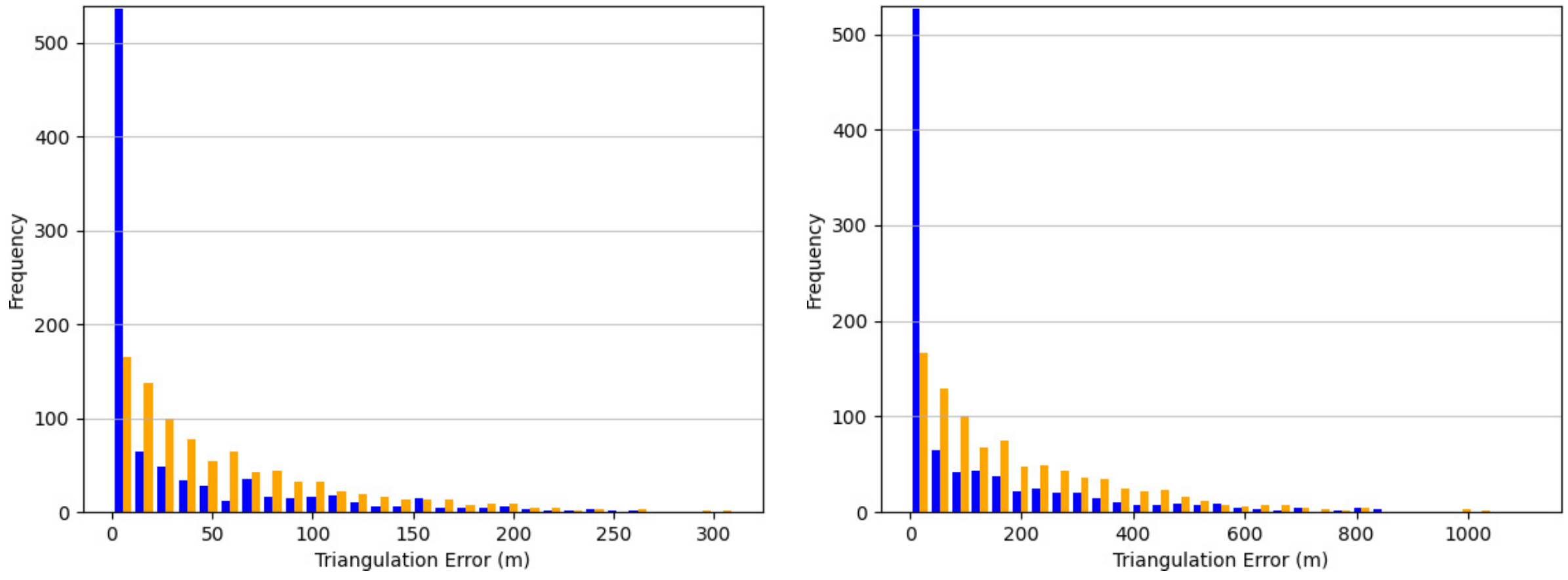

(a) (b)

Fig. 12. (Color online) Histogram of localization errors for a pair of Alcor cameras (a) and ZWO cameras (b). The errors introduced in the simulation were due to the fact that the detected object centroids were erroneous, with an error distribution as described in the text. Note the difference in scale of the $x$-axis in the two graphs.

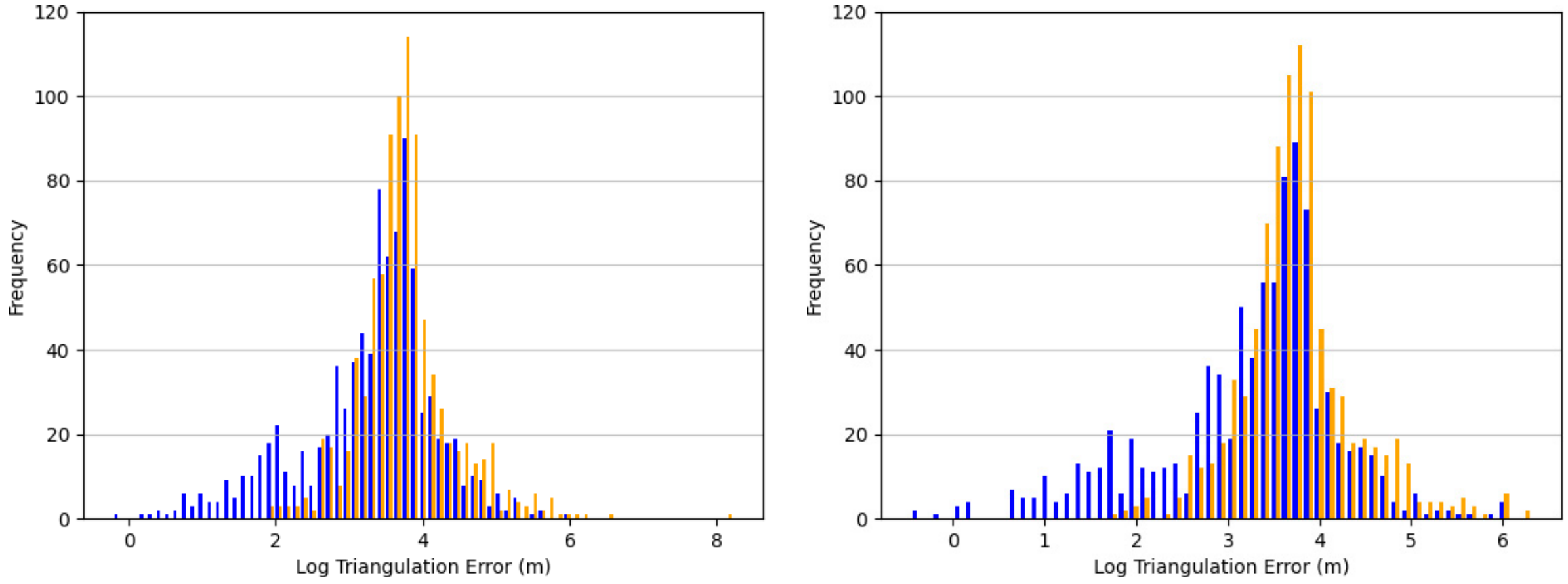

(a) (b)

Fig. 13. (Color online) Histogram of localization errors for a pair of Alcor cameras (a) and ZWO cameras (b). The errors introduced in the simulation were a combination of all error sources described in the text. Note that the $x$-axis has a log (base 10) scale.

Finally, we simulated the situation in which one camera images a nearby object (e.g. an insect) which is too small to be seen by the other cameras. All cameras also image a large, uniformly visible control object, as before, whose position is assigned random values in the range $-2000\,\text{m} \leq x, y \leq 2000\,\text{m}$ and $200\,\text{m} \leq z \leq 2000\,\text{m}$. In none of the 100 simulations is the insect ever reported as localized. The insect is ignored while the control object is localized accurately (with error akin to that reported in Fig. 7).

## 6. Related Work

The hardware and software described in this work are similar to that used to track meteors streaking through the Earth's atmosphere. There exist several long-running meteor- and meteoroid-tracking networks, including:

- The Fireball Recovery and InterPlanetary Observation Network (FRIPON) (Colas *et al.*, 2020)
- The Global Fireball Observatory (GFO) (Devillepoix *et al.*, 2020)
- First Italian Network for Systematic Surveillance of Meteors and Atmosphere (PRISMA) (Gardiol *et al.*, 2016)

- The Global Meteor Network (GMN) (Vida *et al.*, 2021)
- The Spanish Fireball and Meteorite Recovery Network (SPMN) (Peña-Asensio *et al.*, 2021a)
- The Desert Fireball Network (DFN) (Bland *et al.*, 2005)

Howie *et al.* (2017) describe in detail how to construct a meteor-tracking camera network.

The builders of these meteor-tracking networks faced several of the problems we address in this paper, such as

- How to design cameras with sufficient spatial and temporal resolution to meet a project's scientific objectives? (e.g. Howie *et al.* (2017) and Vida *et al.* (2021))
- How to build hardware (e.g. cameras and computers) that can be sited outdoors all year round? (e.g. Howie *et al.* (2017))
- How far should cameras be separated in order to attain a desired object localization accuracy? (e.g. Colas *et al.* (2020))
- How to perform extrinsic calibration for a network of widely-spaced cameras? How often is recalibration required? (e.g. Vida *et al.* (2021), Jeanne *et al.* (2019) and Barghini *et al.* (2019))
- How to compute the trajectory of a moving object as seen simultaneously from multiple widely-spaced cameras? (e.g. Vida *et al.* (2021))
- How to control the timing of image capture among disparate cameras? (e.g. Vida *et al.* (2021))

The above-mentioned camera networks make use of sophisticated object-tracking algorithms. The Global Meteor Network utilizes the method of intersecting planes (Ceplecha, 1987). With this method, a plane is fit to the 3D points comprising the lines of sight from an observer (camera) to a moving meteor over time; at least two such lines of sight are required to form a plane. When two planes are constructed with respect to two different observers viewing the same meteor, the constructed planes intersect at a line; this is the line describing the meteor's trajectory.

Another trajectory-inference method — straight least squares — is described in Borovicka (1990); this is also known as the Line of Sight (LoS) method. With this procedure, each observation of a meteor from each camera is converted into a 3D ray (the so-called LoS for that observation). The 3D line at which all of these lines of sight intersect is considered to be the trajectory of the meteor. Since there is typically measurement error in computing lines of sight, a least-squares fit is used to infer the meteor's trajectory.

A third meteor trajectory solving method is known as the multi-parameter fit (MPF) method (Gural, 2012). Unlike the trajectory solvers described above, MPF attempts to fit values for a meteor's velocity and deceleration as well as its position. MPF utilizes the methods of intersecting planes and LoS to construct an initial guess of a meteor's position over time. The algorithm then tries to fit the meteor's observed velocity and deceleration profile against three different models: constant velocity, velocity with quadratic deceleration as a function of time and velocity with exponential deceleration as a function of time. MPF finds the best set of parameters — including which deceleration profile best fits observations — using an iterative Nelder–Mead algorithm to minimize a suitable cost function.

We note that a difference between meteor trajectory solving and our work is that the former typically assumes a simple object velocity profile (e.g. straight-line motion, constant velocity or quadratic deceleration as a function of time as described above). We cannot assume in this work that the objects we intend to track exhibit simple velocity profiles (e.g. airplanes, helicopters, drones, birds and insects would not have meteoric velocity profiles).

As we described in Sec. 2.1, meteor-tracking camera networks must also calibrate their cameras in order to successfully compute the world coordinate system coordinates of a meteor as seen from disparate cameras pointed in different directions. As these networks often use all-sky cameras, image distortion due to fish-eye lenses must also be corrected by calibration.

Borovicka (1992) provides an influential solution to camera calibration for meteor tracking. The algorithm described in this work assumes that the optical axis of the camera to be calibrated is pointed near the local zenith and the $x$-axis of its image plane is pointing close to due south and that the $y$-axis points close to due west. A relatively simple equation is used to model fish-eye distortion. A camera's distortion parameters and orientation are determined by finding correspondences between stars as seen in an image and those same stars identified in a catalogue (and whose positions in the celestial sphere as a function of time are known from this catalogue); this is a process known as astrometric calibration.

Borovicka *et al.* (1995) improve upon the method described in Borovicka (1992) by introducing a more sophisticated model for fish-eye lens distortion as well as by accounting for the fact that sometimes an imaging plate's normal vector is not perfectly parallel to the optical axis of a camera. Bannister *et al.* (2013) simplify the calibration method put forth in Borovicka *et al.* (1995), and adapt it for charge-coupled device (CCD) cameras.

Jeanne *et al.* (2019) describe a calibration method for cameras in the FRIPON network. Though the FRIPON cameras have fish-eye lenses Jeanne *et al.* (2019) use a software tool called SCAMP (Bertin, 2006) to obtain rough first-order astrometric calibration estimates using the part of the camera image near the image center, where lens distortion is minimal; SCAMP was not designed to account for fish-eye distortion. The authors model fish-eye distortion as a nine-degree polynomial function of $R$, where $R$ is the distance in image coordinates from the image center to an imaged star. Jeanne *et al.* (2019) use the method of Borovicka *et al.* (1995) to account for the fact that the imaging plane normal is not perfectly parallel to the camera's optical axis. These calibration methods bear a strong resemblance to the procedure we document in Sec. 4.2.3; the above references provide much information to guide us in our work.

In addition to meteors, several authors have investigated tracking man-made airborne objects with calibrated cameras. For example, Levit *et al.* (2008) describe tracking the re-entry of the Stardust Sample Return Capsule using a set of visual data taken simultaneously from ground- and aircraft-based cameras. This work differs from ours in that we do not make use of images of objects taken from aircraft for object tracking. Similarly, de Pasquale *et al.* (2009) describe aircraft-based observations of the re-entry of the spacecraft ATV-JV for the purposes of trajectory analysis. Peña-Asensio *et al.* (2021b) report how the Spanish Meteor Network recorded the re-entry trajectory of a Space X Falcon 9 rocket component. This work is relevant in that it describes an extrinsic camera calibration method similar to one we intend to employ (Sec. 4.2.3). The work also relates a method for tracking the Falcon 9 component; as the tracking algorithm assumes the tracked object follows a smoothly curving elliptical path, it is of limited use for our purposes.

Other efforts have been published that use networks of cameras to track both man-made and natural objects in the night sky. For example, Danescu *et al.* (2012) describe a long-baseline (37 km) stereo camera system for tracking night-sky objects (e.g. aircraft, meteors and low-orbit satellites). This work bears some similarity to ours: we both use optical cameras calibrated with the aid of GPS and celestial objects. Our system, though, collects data in non-optical wavelengths as well as employing three independent methods to determine the orientations of our cameras. Cavagna *et al.* (2022) describe a system of stereo optical cameras for localizing Earth-orbiting objects. They provide a comparison of two localization algorithms, and present empirical data regarding sources of localization error for each algorithm; this information will supplement the experiments we describe in Sec. 5.

Peña-Asensio *et al.* (2021c) provide Python software for the automated tracking of meteors given images from widely spaced upward-looking cameras positioned on the ground. This software is used by the Spanish Fireball and Meteorite Network. The software solves several problems which are similar to those we address in this work, including extrinsic calibration of the cameras given identified celestial objects as well as the localization of imaged aerial objects.

## 7. Conclusions and Future Work

In this work, we have described a novel, robust imaging suite that will capture image data in the optical, IR, and NIR wavelengths. The purpose of collecting these data is to further the scientific study of UAPs. In particular in this work, we plan to use these cameras to calculate the location, velocity and acceleration of objects of interest. We have described how these cameras will be calibrated to facilitate accurate localization, and have described the algorithms we will initially employ to solve the correspondence problem and localize objects given image correspondences. We have demonstrated in simulated experiments how our localization estimates will be affected by errors due to miscalibration, differences in image capture times and errors introduced by our object detection algorithm. We have also demonstrated experimentally that our correspondence algorithm will accurately assign image-frame positions to detected objects.

There are a number of directions we intend to take this project in future. For one, the extrinsic calibration method using ADS-B-equipped aircraft described in Sec. 4.2.1 is novel. We intend to describe this algorithm along with the experimental assessment of the algorithm in a stand-alone paper in future. Also, the extrinsic calibration methods described in Secs. 4.2.1 and 4.2.2 required manual work to identify aircraft and drones in images; we will attempt to automate this identification.

We demonstrated in Sec. 5 that non-synchronized image capture introduces non-negligible inaccuracy in object localization. One way to mitigate this source of error could be to improve time synchronization across the computers to which the cameras are attached. Mkacher and Duda (2019) describes a method for improving synchronization accuracy using NTP, which we will explore in future.

There are numerous alternatives to the triangulation algorithm we outlined in Sec. 4.5.1. As we demonstrated in Sec. 5, this algorithm contributes to our localization error. We intend to evaluate alternatives to the LTM-H algorithm that might have superior performance. This list of alternatives includes but is not limited to: Statistical Angular Error-Based Triangulation (Recker *et al.*, 2013), Multi-View Stereo (Rumpler *et al.*, 2011), and Three-Camera Maximum Likelihood Estimation (Torr and Zisserman, 1997).

As described in Sec. 6, Peña-Asensio *et al.* (2021c) provide Python code for both camera calibration and trajectory tracking of meteors. It will be useful to compare their solutions to ours and to perhaps adapt for our purposes the open source code that Peña-Asensio *et al.* (2021c) provide.

## Appendix A. Analysis of Dalek Cooling System

In order to determine how much heat must be expelled from the Dalek's dome, we must ensure that the following sum is less than 0:

$$Q(\text{sun}) + Q(\text{cameras}) + Q(\text{fan}) - Q(\text{air}), \quad \text{(A.1)}$$

where

- $Q(\text{sun}) = 1000 \times 0.2 \times 0.84 = 168\,\text{W}$ since the dome has surface area $0.2\,\text{m}^2$; solar irradiance is $1000\,\text{W/m}^2$; and the dome's emissivity is 0.84.
- $Q(\text{cameras}) = 5\,\text{W}$ as each camera produces approximately 500 mW of heat.
- $Q(\text{fan}) = 1.6\,\text{W}$; this is the input heat from the fan.
- Given our fan's speed and size and the area of the dome's air inlet and outlet filters, we expect $Q(\text{air})$ to be about 277 W.

Thus $Q(\text{sun}) + Q(\text{cameras}) + Q(\text{fan}) - Q(\text{air}) = 168\,\text{W} + 5\text{W} + 1.6\,\text{W} - 277\,\text{W} = -102\,\text{W}$ which is indeed less than 0. In this analysis, we have neglected external air convection to simulate a worst-case scenario.

## Appendix B. Analysis of Distance Estimation Error Due to Unsynchronized Cameras Imaging a Moving Object

Here, we investigate the error in object distance estimation due to image capture time mis-synchronization between two cameras. This error analysis is an extension of that done in Sec. 2.2.

We formulate the problem as follows: our 3D world coordinate system is defined such that its origin is at a point labeled $L$. There is another point $R$ at location $(b, 0, 0)$; thus the segment from $L$ to $R$ is parallel to the $x$-axis of the world coordinate system. There is a third point $O$ at $(o_x, o_y, o_z)$; the $z$-axis is perpendicular to the $x$-axis and intersects with $O$. The $y$-axis is perpendicular to both the $x$- and $z$-axes. Thus, $o_y$ is 0. The coordinate system is right-handed.

Our two cameras are located at $L$ and $R$, respectively. They are oriented identically, such that their optical axes are parallel to the $z$-axis of the world coordinate system and the $x$- and $y$-axes of their imaging planes are parallel to the $x$- and $y$-axes (respectively) of the world coordinate system.

At time $t = 0$, there is an object-of-interest positioned at $O$. The distance to be measured–$a$–is the distance from $L$ to $O$ at time $t = 0$. We define the angle $\theta$ as the angle between the LoS vector from $L$ to $O$ and the $x$-axis. The angle $\alpha = 180 - \theta$.

After the object is imaged by the camera at $L$, it moves for a duration $\Delta t$ time units in the direction $V = (v_x, v_y, v_z)$. Thus, the object moves by $V\Delta t$ distance units, ending up at the position $O + V\Delta t$. The object is then imaged by the camera at $R$. The angle between the LoS vector from $R$ to $O + V\Delta t$ and the $x$-axis is defined as $\beta$.

The formulation is very similar to that given depicted in Fig. 2. The difference is that in this analysis we allow each element of $V$ to be nonzero;

in the analysis of Sec. 2.2, only $v_x$ was allowed to be nonzero.

We use Eq. (4) to compute the apparent distance from $L$ to $O$. When $V\Delta t = (0,0,0)$, Eq. (4) will yield the correct distance from $L$ to $O$; that is, $Z$ will equal $a$. When $V\Delta t \neq (0,0,0)$, $Z$ will typically differ from $a$; we shall label this distance $Z_{\Delta t}$ and we shall label this difference as $c$ (i.e. $c = Z_{\Delta t} - a$).

We shall assume that $o_z > 0$ and $o_z + v_z\Delta t > 0$. This assumption ensures that the imaged object remains in sight of both cameras at times $t = 0$ and $t = \Delta t$.

From the above it follows that

$$\alpha = \pi - \operatorname{atan}\left(\frac{o_z}{o_x}\right). \tag{B.1}$$

At time $t = 0$, it is the case that

$$\beta_0 = \operatorname{acos}\left(-\frac{-b + o_x}{\sqrt{o_z^2 + (b - o_x)^2}}\right). \tag{B.2}$$

At time $t = \Delta t$

$$\beta_{\Delta t} = \operatorname{acos}\left(-\frac{-b + o_x + v_x\Delta t}{\sqrt{v_y\Delta t^2 + (-o_z - v_z\Delta t)^2 + (b - o_x - v_x\Delta t)^2}}\right). \tag{B.3}$$

By substituting Eqs. (B.1) and (B.2) into Eq. (4) and simplifying, we get

$$a = \frac{b\sqrt{-\frac{(-b+o_x)^2}{o_z^2+(b-o_x)^2} + 1}}{\sin\left(\operatorname{acos}\left(-\frac{-b+o_x}{\sqrt{o_z^2+(b-o_x)^2}}\right) + \operatorname{atan}\left(\frac{o_z}{o_x}\right)\right)}. \tag{B.4}$$

Using the same procedure for $Z_{\Delta t}$, we get

$$Z_{\Delta t} = \frac{b\sqrt{-\frac{(-b+o_x+v_x\Delta t)^2}{v_y\Delta t^2+(-o_z-v_z\Delta t)^2+(b-o_x-v_x\Delta t)^2} + 1}}{\sin\left(\operatorname{acos}\left(-\frac{-b+o_x+v_x\Delta t}{\sqrt{v_y\Delta t^2 + (-o_z - v_z\Delta t)^2 + (b - o_x - v_x\Delta t)^2}}\right) + \operatorname{atan}\left(\frac{o_z}{o_x}\right)\right)} \tag{B.5}$$

Since $c = Z_{\Delta t} - a$

$$c = -\frac{b\sqrt{-\frac{(-b+o_x)^2}{o_z^2+(b-o_x)^2} + 1}}{\sin\left(\operatorname{acos}\left(-\frac{-b+o_x}{\sqrt{o_z^2+(b-o_x)^2}}\right) + \operatorname{atan}\left(\frac{o_z}{o_x}\right)\right)} + \frac{b\sqrt{-\frac{(-b+o_x+v_x\Delta t)^2}{v_y\Delta t^2+(-o_z-v_z\Delta t)^2+(b-o_x-v_x\Delta t)^2}+1}}{\sin\left(\operatorname{acos}\left(-\frac{-b+o_x+v_x\Delta t}{\sqrt{v_y\Delta t^2 + (-o_z - v_z\Delta t)^2 + (b - o_x - v_x\Delta t)^2}}\right) + \operatorname{atan}\left(\frac{o_z}{o_x}\right)\right)} \tag{B.6}$$

Now we explore the behavior of $\frac{c}{a}$ as was done in Sec. 2.2. With $v_y$ and $v_z$ set to 0, $\frac{c}{a}$ (i.e. Eq. (B.6) divided by Eq. (B.4) becomes

$$\frac{c}{a} = \frac{o_z v_x\Delta t}{bo_z - o_x o_z + o_x|o_z| - o_z v_x\Delta t} \tag{B.7}$$

Since $o_z > 0$, we can replace $|o_z|$ with $o_z$ and Eq. (B.7) becomes

$$\frac{c}{a} = \frac{v_x\Delta t}{b - v_x\Delta t} \tag{B.8}$$

which is exactly Eq. (8).

With $v_x$ and $v_z$ set to 0, $\frac{c}{a}$ becomes

$$\frac{c}{a} = \frac{o_z(b\sqrt{o_z^2 + v_y\Delta t^2} - b|o_z| - o_x\sqrt{o_z^2 + v_y\Delta t^2} + o_x|o_z|)}{(bo_z - o_x o_z + o_x\sqrt{o_z^2 + v_y\Delta t^2})|o_z|} \tag{B.9}$$

Since $o_z > 0$, we can replace $|o_z|$ with $o_z$. Equation (B.9) can be simplified to

$$\frac{c}{a} = \frac{-bo_z + b\sqrt{o_z^2 + v_y\Delta t^2} + o_x o_z - o_x\sqrt{o_z^2 + v_y\Delta t^2}}{bo_z - o_x o_z + o_x\sqrt{o_z^2 + v_y\Delta t^2}} \tag{B.10}$$

In this case, the error in the distance estimate grows large when the denominator of Eq. (B.11) is close to 0 or in other words when $v_y\Delta t \approx \pm\sqrt{b}\sqrt{b - 2o_x\frac{o_z}{o_x}}$.

With $v_x$ and $v_y$ set to 0, $\frac{c}{a}$ becomes

$$\frac{c}{a} = \frac{o_z(-b|o_z| + b|o_z + v_z\Delta t| + o_x|o_z| - o_x|o_z + v_z\Delta t|)}{(bo_z - o_x o_z + o_x|o_z + v_z\Delta t|)|o_z|}. \tag{B.11}$$

Since $o_z > 0$, we can replace $|o_z|$ with $o_z$. Also since $o_z + v_z\Delta t > 0$ we can replace $|o_z + v_z\Delta t|$ with $o_z + v_z\Delta t$. Equation (B.11) then becomes

$$\frac{c}{a} = \frac{v_z\Delta t(b - o_x)}{bo_z + o_x v_z\Delta t}. \tag{B.12}$$

In this case, the error in the distance estimate grows large when $bo_z + o_x v_z\Delta t$ is close to 0 or in other words when $v_z \Delta t \approx \frac{-bo_z}{o_x}$. For example, when $\frac{b}{o_x}$ is close to 1 then a move by $-o_z$ units in the $z$-direction (i.e. to the baseline) will produce a relatively large error in the distance estimate.